\newcommand{\ket}[1]{\lvert #1 \rangle}
\newcommand{\bra}[1]{\langle #1 \lvert}
\newcommand{\dif}{\text{d}}

\newcommand{\Lagr}{\mathcal{L}}
\newcommand{\fourint}{\int \frac{\dif^4 k}{(2\pi)^4}\,}

\newcommand{\Tr}{\text{Tr}}

\newcommand{\detM}{\text{det}}
\newcommand{\kond}[2]{\langle \bar #1 #2 #1 \rangle}

\documentclass[%
amsmath,amssymb,
reprint,%
preprintnumbers,%
nofootinbib
]{revtex4-2}

\usepackage{graphicx}
\usepackage{dcolumn}
\usepackage{bm}

\usepackage{simplewick}
\usepackage{slashed}
\usepackage{dsfont}
\usepackage{amsthm}
\usepackage{mathtools}
\usepackage{colortbl} 
\usepackage{xcolor}

\usepackage{lipsum}
\usepackage{multirow}
\usepackage{bigdelim}
\usepackage[%
	bookmarksnumbered,
	bookmarksdepth=5,
	colorlinks=false,
	linkbordercolor=red%
	]{hyperref}

\begin{document}
	
	\preprint{MITP-22-003}
	\preprint{\today}
	
	\title[Sample title]{Impact of unphysical meson decays on the parameter fixing\\ in the three-flavour Nambu-Jona-Lasinio model\\ for different regularisation methods}
	\author{Dominic Kraatz}
	\email{kraatz@itp.uni-frankfurt.de}
	\affiliation{
		Institute for Theoretical Physics,\\ Johann Wolfgang Goethe-Universit\"at, 60438 Frankfurt am Main, Germany}
	
	\author{Michel Stillger}%
	\email{m.stillger@uni-mainz.de}
	\affiliation{PRISMA$^+$ Cluster of Excellence \& 
		Mainz Institute for Theoretical Physics,\\ Johannes Gutenberg University, 55099 Mainz, Germany
		\\
	}%
	
	
	\begin{abstract}
		We investigate the influence of the allowed unphysical meson decay into a quark-antiquark pair of the three flavour Nambu--Jona-Lasinio (NJL) model on its parameter fixing. This decay manifests in a non-vanishing imaginary part of the corresponding meson propagator and its polarisation loop, respectively. In order to handle the emerging divergent integrals, we focus on the Pauli-Villars, three- and four momentum cutoff regularisation method. Here, we fix the parameters to certain observables, which are effected by the non-vanishing imaginary part. The resulting parameter-sets for each regularisation scheme are compared to each other with and without taking the imaginary part of the polarisation loop into account.
	\end{abstract}
	
	\pacs{Valid PACS appear here}
	\keywords{Suggested keywords}
	\maketitle
	
	
	\section{Introduction}
	
	The NJL model is a well know and often used model to investigate the properties of stongly interacting matter in vacuum as well as under extreme conditions, i.e. in hot and dense matter. Originally the model was designed by Y. Nambu and G. Jona-Lasinio in 1961 to explain the high nucleon mass in accordance with the partially conserved axial current in the pre-QCD era~\cite{PhysRev.122.345, PhysRev.124.246}. Later, it was reinterpreted for QCD calculations in the low-energy regime, where gluon degrees of freedom are supposed to be frozen-out. This leads to a reduction of the complex structure of strong interactions in the QCD Lagrangian to a point-like local interaction. In particular, the NJL model shares the same global symmetries with QCD and in this way also can describe the chiral symmetry and its spontaneous breaking in the vacuum as well. Moreover, the NJL model is often used to study the thermodynamic properties of QCD like the restoration of the chiral symmetry for high temperatures and densities in the QCD phase diagram~\cite{njl-buballa, oertel2000investigation, BraunMunzinger:2008tz, njl-klevansky}, and has even been used to investigate baryons, cf.~\cite{Buck:1992, Torres-Rincon:2015rma}.\\
	
	Due to the non-renormalizable character of the model, an additional cutoff parameter $\Lambda$ appears, which has to be included in order to handle the upcoming divergent integrals. Hereby, different methods can be used like: (1) Introducing a sharp three-momentum or four-momentum cutoff parameter. (2) Introducing additional \mbox{(counter-)terms} to control the divergences of the integrals for high momenta. After choosing an appropriate regularisation method the other parameters of the model can be fixed to physical observables. In our case we fix to the masses of pion, kaon, $\eta$- and $\eta'$-meson as well as to the pion decay constant \cite{njl-klevansky,njl-buballa,Kohyama:2016fif}.\\
	
	Besides the non-renormalizable character of the NJL model, another drawback is the lack of confinement due to the missing gluonic interactions. While some thermodynamic properties of QCD in the medium can be simulated by including Polyakov loops ~\cite{PhysRevD.73.014019, Blanquier:2016dls}, the meson decay into a quark-antiquark-pair is still allowed and possible. By Choosing heavy enough up- and down-quark masses, these decays can be circumvented for pions and kaons. However, for reasonable quark masses the decay is still possible for heavier mesons like the $\eta'$-meson. This results in non-vanishing imaginary parts of polarisations loops and, therefore, in complex-valued meson propagators.\\
	
	In this paper we create and compare parameter sets for the three-flavour NJL model using the common regularisation methods mentioned earlier. Hereby, we focus on the treatment of the emerging imaginary part of the meson propagator, which has not been studied in detail so far. In order to understand the contribution of the unphysical property of the model to allow decays into quark-antiquark pairs, we take a look at parameter sets with and without taking the imaginary part into account. However, other authors tend to use parameter sets from older publications, like~\cite{HATSUDA1994221,Rehberg:1995kh}, or create their own ones with limited information of the used fit parameters or handle of imaginary parts. \\
	
	This paper is organised as follows:
	At first we want to introduce the NJL Lagrangian and shortly review the used methods in order to investigate the quark and mesonic properties of the model. Then, the method for the parameter creation for the different regularisation schemes is introduced. Hereby, we discuss the effect of the decay into a quark-antiquark pair as well as other emerging critical phenomena for the model. Of course, the parameter sets are presented and compared to each other.
	
	We want to mention at this point that the NJL model in general have often been discussed in the literature and therefore we are not going into much detail. For further information we refer as an example to~\cite{njl-buballa, njl-klevansky, Klimt:1989pm, Vogl:1989ea, Alkofer:295076, Reinhardt:1989rw} and the literature mentioned in these works as well. 
	
	\section{The three-flavour NJL model}
	In this paper we consider the following NJL Lagrangian:
	\begin{equation}
		\Lagr^{\text{NJL}}=\bar \psi\left(i\slashed{\partial}-m_0\right)\psi+\Lagr^{(4)}+\Lagr^{(6)} \label{equ:njl_lagr_basic}
	\end{equation}
	A NJL-type Lagrangian can be in general separated into two parts: The first one is the interaction-free part leading to the Dirac equation, which describes the kinematic of the system, while the second one represents the point-like interaction within the model.\footnote{The three relevant spaces are ordered as \mbox{Dirac $\otimes$ Colour $\otimes$ Flavour}}

	In our case with $N_f=3$ flavours, the four-point interaction term is given by
	\begin{equation}
		\Lagr^{(4)}=G\sum_{a=0}^{8}\left[(\bar\psi\tau_a \psi)^2+(\bar\psi i\gamma_5\tau_a \psi)^2\right]\,, \label{equ:njl_four_point}
	\end{equation}
	with $\tau_a$ denoting the eight Gell-Mann-matrices of the SU(3) plus $\tau_0=\sqrt{2/N_f}\mathds{1}$.	
	
	The latter term in equation~\eqref{equ:njl_lagr_basic} represents an instanton induced six-point interaction Lagrangian ('t Hooft term) in order to break the otherwise remaining $U_{\text{A}}(1)$ symmetry in the three flavour case explicitly. It is given by a determinate over flavour space~\cite{Osipov:2004mn}
	\begin{equation}
		\Lagr^{(6)}=K\left[\detM_f\left(\bar\psi(1+\gamma_5)\psi\right)+\detM_f\left(\bar\psi(1-\gamma_5)\psi\right)\right]\label{equ:njl_six_point}\,.
	\end{equation}
	As one can see directly, the six-point interaction Lagrangian leads to a mixing of different quark flavours. We want to emphasise at this point that the upper given term is not unique and other interaction terms will lead also to an explicit breaking of the $U_{\text{A}}(1)$ symmetry. However, equation~\eqref{equ:njl_six_point} is the commonly used term.
	
	In the upper equations, $\psi$ denotes a Dirac spinor, representing a quark field. Additional indices for the colour and flavour space are for now omitted to increase readability. The \mbox{mass $m_0$} denotes a diagonal matrix of dimension $N_f$ containing the different current (or bare) masses of the quarks.

	\subsection{Quark properties}
 	It is now convenient to introduce the quark condensate $\kond{\psi_f}{}$ for a certain flavour $f$ which is given by
	\begin{equation} \label{equ:quark_condensate}
		\begin{split}
			\kond{\psi_f}{} &= - \fourint \Tr\, S_f(k) \\
			&= - 4 N_c \, m_f \, I_1(m_f) \, ,
		\end{split}
	\end{equation}
	where $S_f$ denotes the quark propagator of flavour $f$ and basically describes a closed quark-loop. A detailed discussion of the emerging integral $I_1$ can be found in appendix~\ref{app:integrals}.
	
	Note that in our approach non-scalar condensates and flavour-mixing ones vanish due to the trace in all three spaces.
	At first, the quark properties of the model is investigated in the mean-field approach by applying
	\begin{equation}
		\left(\bar\psi\Gamma_I\psi\right)\approx\kond{\psi}{\Gamma_I}+\delta_{\Gamma_I}
	\end{equation}
	to the interaction Lagrangian's~\eqref{equ:njl_four_point} and~\eqref{equ:njl_six_point} and neglecting terms of order two in $\delta_{\Gamma_I}$. This leads to a linearisation of both Lagrangian's such that the coupled gap-equations for a certain flavour can be read off directly
	\begin{equation}
		m_i=m_{0,i}-4G\kond{\psi_i}{}+2K\kond{\psi_j}{}\kond{\psi_k}{}
	\end{equation}
	with $i\neq j \neq k \neq i \in \{u, d, s\}$ and $m_f$ denoting the so-called constituent quark mass.\footnote{A more formal way is to perform a Hubbard-Stratonovich transformation. However, at the end, both approaches lead to the same result.}\\	
	In case of the isospin limit ($m_u=m_d$), which will always be applied in this work, the upper equation further simplifies to
	\begin{equation} \label{equ:gapequ}
		\begin{split}
			m_u &= m_{0,u} -4G\kond{\psi_u}{}+2K\kond{\psi_u}{}\kond{\psi_s}{}\\
			m_s &= m_{0,s} -4G\kond{\psi_s}{}+2K\kond{\psi_u}{}\kond{\psi_u}{}\,.
		\end{split}
	\end{equation}
	
	\section{Mesonic spectrum}
	\subsection{Bethe-Salpeter equation and mesonic propagators} \label{subsec:meson_properties}
	\begin{figure}[h]
		\centering
		\includegraphics[width=0.48\textwidth]{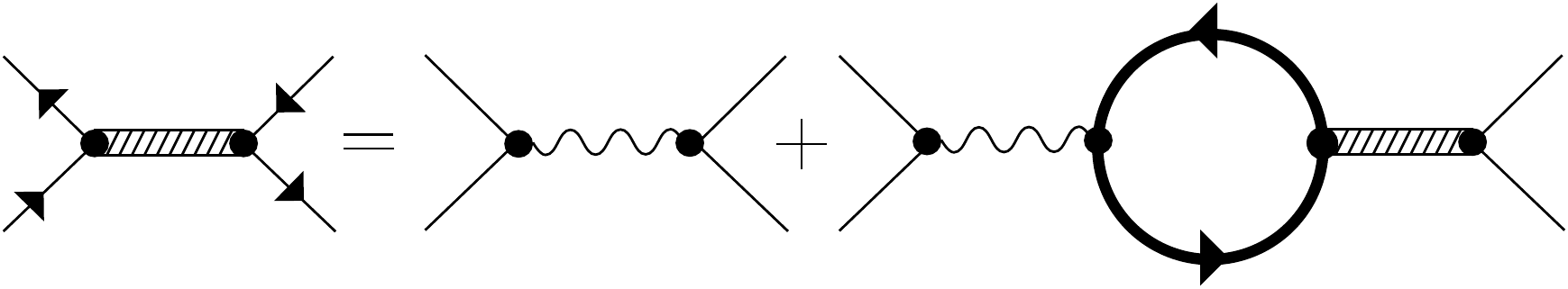}
		\caption{\label{fig:MesonBSE-SC} Self-consistent expression for the Bethe-Salpeter equation in RPA.}
	\end{figure}	
	
	To obtain an expression for the meson propagator, and thus for the properties of the mesons, the self-consistent Bethe-Salpeter equation (BSE) in random-phase approximation (RPA), 
	\begin{equation}
		i\mathcal{T}(p)=i\mathcal{K}+ i\mathcal{K}(-i\Pi(p))i\mathcal{T}(p)\label{equ:BSE}\,,
	\end{equation}
	diagrammatically represented in FIG.~\ref{fig:MesonBSE-SC} has to be solved. Here, $\mathcal{K}$ and $\Pi$ denote the scattering kernel and the polarisation loop, respectively. $\mathcal{T}$ denotes the so-called scattering matrix (T-matrix).

	In random-phase approximation, where only quark-antiquark polarisation loops are included, the right-hand side of the BSE is the sum over all these loops. The leading order term therefore is the scattering kernel itself.

	As one can see immediately from the upper description of the BSE, it would be sufficient to have a four-point like interaction in order to evaluate the diagrams within FIG.~\ref{fig:MesonBSE-SC} more or less directly:
	
	In case of the t'Hooft Lagrangian~\eqref{equ:njl_six_point} it is possible to reduce its six-point character to a four-point one by closing certain quark loops. For further details on the technical procedure to obtain a four-point like interaction out of the six-point one, we refer to~\cite{njl-klevansky}. However, the original four-point interaction and the new one can be combined to an effective four-point interaction Lagrangian 
	\begin{widetext}
		\begin{equation}\label{equ:eff-four-point_L}
			\begin{split}
				\Lagr^{(4)}_\text{eff}=\sum_{a=0}^8\left(G^{(-)}_a(\bar\psi\tau_a\psi)^2+G^{(+)}_a(\bar\psi i\gamma_5\tau_a\psi)^2\right)&+\left[G^{(-)}_{80}(\bar\psi\tau_8\psi)(\bar\psi\tau_0\psi)+G^{(+)}_{80}(\bar\psi i\gamma_5\tau_8\psi)(\bar\psi i\gamma_5\tau_0\psi)\right]\\
				&+\left[G^{(-)}_{08}(\bar\psi\tau_0\psi)(\bar\psi\tau_8\psi)+G^{(+)}_{08}(\bar\psi i\gamma_5\tau_0\psi)(\bar\psi i\gamma_5\tau_8\psi)\right]\,,		
			\end{split}		
		\end{equation}
	\end{widetext}	
	where the (redefined) flavour dependent coupling constant $G_a^{\pm}$ can be expressed in terms of the quark condensate
	\begin{equation}
	\begin{split}
	&G^{(\pm)}_0=G\pm\frac{1}{3}K(\kond{\psi_s}{}+2\kond{\psi_u}{})\\
	&G^{(\pm)}_1=G^{(\pm)}_2=G^{(\pm)}_3=G\mp \frac{1}{2}K\kond{\psi_s}{}\\
	&G^{(\pm)}_4=G^{(\pm)}_5=G^{(\pm)}_6=G^{(\pm)}_7=G\mp \frac{1}{2}K\kond{\psi_u}{}\\
	&G^{(\pm)}_8=G\pm\frac{1}{6}K(\kond{\psi_s}{}-4\kond{\psi_u}{})\,.
	\end{split}\label{equ:coulings_3_flavour}	
	\end{equation}
	Since there is no flavour mixing contribution in the original four-point interaction Lagrangian, the terms related to this mixing comes with the effective coupling strength
	\begin{equation}
	G^{(\pm)}_{80,08}=\pm\frac{1}{6}K\sqrt{2}(\kond{\psi_s}{}-\kond{\psi_u}{})\,,\label{equ:coupling_eta}
	\end{equation}
	which only depends on the coupling $K$ of the t'Hooft determinant.  
	Based on our assumption of degenerate up- and down-quark masses (isospin limit), we have no term in the Lagrangian that includes a mixing of $\tau_3-\tau_8$ or $\tau_0$, respectively. Hence, a mixing of the $\pi$ and $\eta$ modes does not appear. Moreover, this assumption leads to a degeneration of each of the three pions and four kaons as well.
	
	Coming back to the BSE, the scattering kernel for the pseudo-scalar mesons, summarized in TABLE~\ref{tab:scat_kernel_pol_loop_3fl}, are given by the effective interaction Lagrangian~\eqref{equ:eff-four-point_L}. The polarisation loop for a certain meson is given by
	\begin{widetext}
		\begin{equation} \label{equ:polarisation_loops}
		-i\Pi^{\text{N}}_{ab}(p)\equiv-\fourint\Tr\left[\Gamma^{\text{N}}\tau_a S(p+k)\Gamma^{\text{N}} \tau_b^{\dagger} S(k)\right]\,.
		\end{equation}		
	\end{widetext}
	Here, $\Gamma^{\text{N}}$ denotes the structure of the corresponding vertex in colour and Dirac-space and is given by 
	\begin{equation}
		\Gamma^{\text{PS}}=i\gamma_5\otimes \mathds{1}_c\,,
	\end{equation}
	where PS denotes the pseudo-scalar channel.
	\begin{table}
		\caption{\label{tab:scat_kernel_pol_loop_3fl} Scattering kernel and corresponding Gell-Mann matrices for pseudo-scalar mesons.}
		\begin{ruledtabular}
				\begin{tabular}{l|ccc}
						meson& $i\mathcal{K}_{ab}^{\text{PS}}$&$\tau_a$&$\tau_b^{\dagger}$\\
						\hline
						&&&\\
						$\ket{\pi^0}$&$2G_{3}^{(+)}$&$\tau_3$&$\tau_3$\\
						$\ket{\pi^{\pm}}$&2$G_{3}^{(+)}$&$\frac{1}{\sqrt{2}}\left(\tau_1+ i\tau_2\right)$&$\frac{1}{\sqrt{2}}\left(\tau_1- i\tau_2\right)$\\
						$\ket{K_0},\ket{\bar{K}_0}$&$2G_{4}^{(+)}$&$\frac{1}{\sqrt{2}}\left(\tau_6+ i\tau_7\right)$&$\frac{1}{\sqrt{2}}\left(\tau_6- i\tau_7\right)$\\
						$\ket{K_{\pm}}$&$2G_{4}^{(+)}$&$\frac{1}{\sqrt{2}}\left(\tau_4+ i\tau_5\right)$&$\frac{1}{\sqrt{2}}\left(\tau_4- i\tau_5\right)$\\
						\ldelim\{{4}{11pt}[$\ket{\eta},\ket{\eta'}$]& $2G_0^{(+)}$&$\tau_0$&$\tau_0$\\
						& $2G_8^{(+)}$&$\tau_8$&$\tau_8$\\
						& $2G_{08}^{(+)}$ & $\tau_0$ & $\tau_8$ \\
						& $2G_{80}^{(+)}$ & $\tau_8$ & $\tau_0$ \\
					\end{tabular}
			\end{ruledtabular}
		\end{table}
	Explicit expressions for the relevant polarisation loops can be found in appendix~\ref{app:polarisation_loops}. Other combinations lead to a vanishing polarisation loop and therefore no bound-states can be obtained. 
	
	However, the right-hand side of the BSE gives rise to a self-consistent scalar function which can be interpreted as a meson propagator of a certain mode. However, in the vicinity of the pole, this function is expected to behave like a free scalar-particle propagator of mass $m_M$. Hence, the left-hand side of \eqref{equ:BSE} can be written as
	\begin{equation} \label{equ:BSE_meson_exchange}
		i\mathcal{T}_{ab}(p) = \left(-i g_{M\bar{\psi}\psi}\right) \frac{i}{p^2-m_M^2} \left(-i g_{M\bar{\psi}\psi}\right) \delta_{ab} \, ,
	\end{equation}
	where $g_{M\bar{\psi}\psi}$ denotes the so-called meson-antiquark-quark coupling. In the context of the BSE as a description of a quark-antiquark scattering process, we can compare both sides near the pole and obtain 
	\begin{equation} \label{equ:meson_quark_antiquark_coupling}
		g_{M \bar{\psi} \psi}^{-2}=\left.\frac{\partial \Pi_M(p^2)}{\partial p^2}\right|_{p^2=m_M^2}
	\end{equation}
	for a certain meson $M$. We want to emphasise at this point, that the meson-antiquark-quark coupling in this work will only be evaluated for the pion because it arises in the description of the later introduced pion-decay constant, cf.~\ref{subsec:pdc}. In the $\eta$-$\eta^{\prime}$-channel equation~\eqref{equ:BSE_meson_exchange} is more complicated due to the off-diagonal structure~\eqref{equ:eta_matrices}.
	
	Based on the equations~\eqref{equ:BSE} and~\eqref{equ:BSE_meson_exchange}, the meson pole masses can be obtained by solving the eigenvalue problem
	\begin{equation}
		\detM\left(\mathds{1}-\mathcal{K} \, \Pi(p^2)\right)=0 \label{equ:det_equation}
	\end{equation}  

	For the pion and kaon, the scattering kernels and polarisation loops can be described by fully diagonal matrices and hence can be treated in a simple way,
	\begin{align} 
		\label{equ:pion_prop}
		D_{\pi}^{-1}(p^2) &\equiv 1-2 G_3^{(+)} \Pi_{\pi}(p^2) \, ,
		\\
		\label{equ:kaon_prop}
		D_K^{-1}(p^2) &\equiv 1- 2 G_4^{(+)} \Pi_K(p^2) \, .
	\end{align}
	 Meanwhile, the $\eta$-meson states are mixed states of the $\eta_0$- and $\eta_8$-meson with a particular mixing angle in nature. We also see this kind of behaviour in the effective Lagrangian~\eqref{equ:eff-four-point_L}. Hence, the scattering kernel as well as the polarisation loop in the flavour channel obtain non-diagonal elements 
	\begin{equation} \label{equ:eta_matrices}
		\mathcal{K}_{ab}=\begin{pmatrix}
		\mathcal{K}_{00}&\mathcal{K}_{08}\\
		\mathcal{K}_{80}&\mathcal{K}_{88}\\
		\end{pmatrix}\,,\quad
		\Pi_{ab}=\begin{pmatrix}
		\Pi_{00}&\Pi_{08}\\
		\Pi_{80}&\Pi_{88}\\
		\end{pmatrix}\,,
	\end{equation} 
	where we dropped the superscript PS for the sake of readability. It follows, that for the two \mbox{$\eta$-mesons} the matrix structure of equation~\eqref{equ:BSE} has to be taken into account. The scattering kernel in terms of the coupling constants are given in TABLE~\ref{tab:scat_kernel_pol_loop_3fl}.\\
	
	Overall, the matrix part within the determinant of equation~\eqref{equ:det_equation} becomes block-diagonal with one $7\times7$ diagonal matrix (three pions, four kaons) and the one $2\times2$ matrix, defined above, for the two $\eta$-mesons.\\

	\subsection{$\eta$-propagator}\label{subsec:meson_prop}
\begin{figure}[t]
		\centering
		\includegraphics[width=0.45\textwidth]{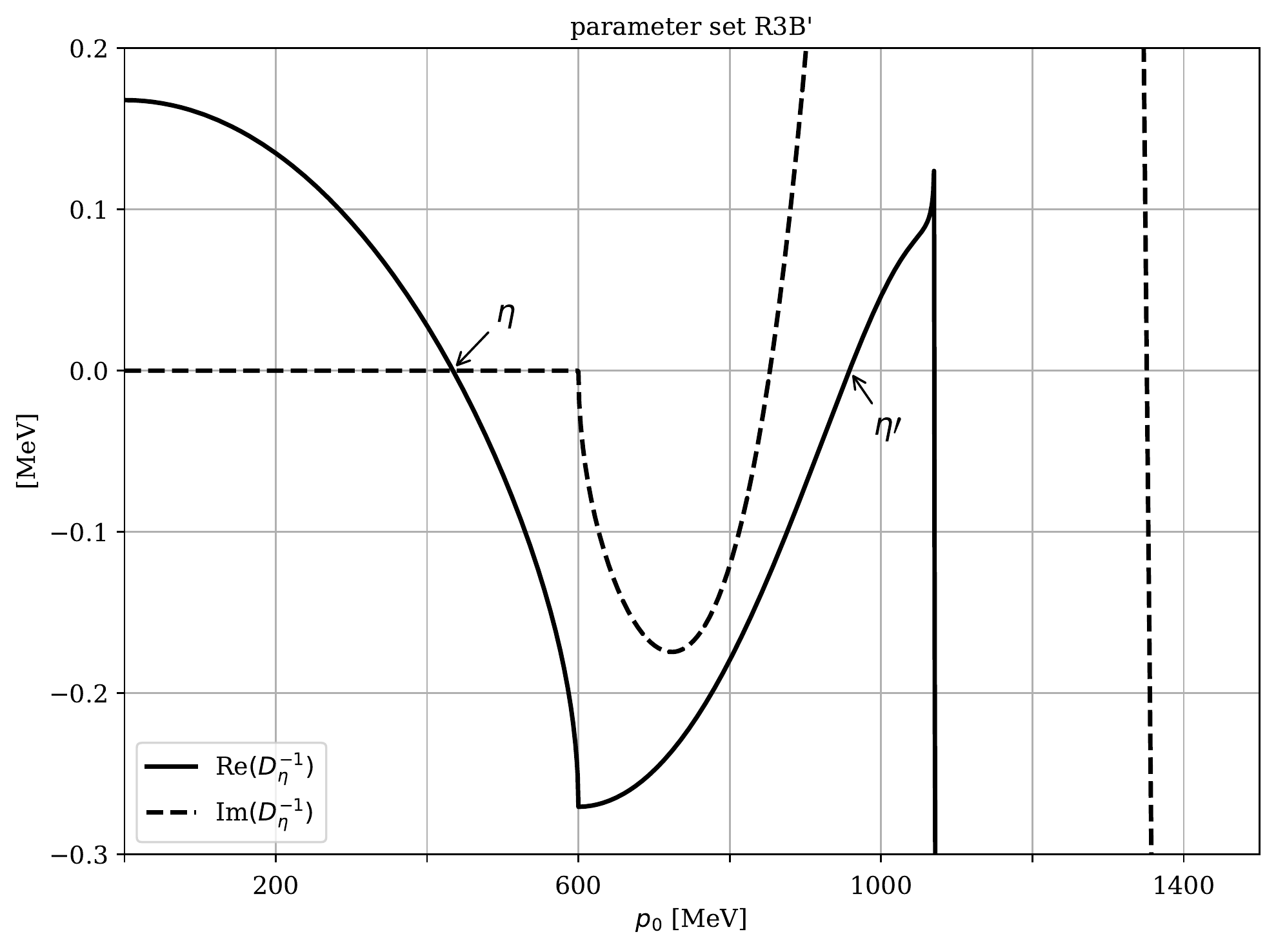}
		\caption{\label{fig:eta_prop_} Real and imaginary part of the inverse $\eta$-propagator as function of $p_0$ (parameter set R3B', cf. TABLE~\ref{tab:parameter_set_PV3R_etaprime}). First and second root of real part are related to the mass of the $\eta$- and $\eta'$-mass, respectively.}
	\end{figure}

	Before we want to start introducing the different regularisation methods, we want to review the properties and behaviour of the meson propagators in more detail. Therefore, we will use the parameters of the Pauli-Villars regularisation scheme with three regulators. However, the discussion in this subsection is not restricted to this method only but can be applied to all others as well.\\
	We have already seen in the previous sections that the $\eta$-propagator will be of particular interest since it gives rise to the $\eta$ and $\eta'$ mass as well.\footnote{We will call the $\eta$ / $\eta'$-meson propagator only $\eta$-propagator.} Hence, we will choose the $\eta$-propagator as an example in order to show the general analytic properties of the propagators emerging in this work.
	After evaluating the determinant of the non-diagonal block-matrix in equation~\eqref{equ:det_equation} which is related to $\eta_0$ and $\eta_8$ we find that the inverse $\eta$-propagator is given by
	\begin{widetext}
		\begin{equation} \label{equ:eta_prop}
			D_{\eta}^{-1}(p^2)\equiv	(1-\mathcal{K}_{00}\Pi_{00})(1-\mathcal{K}_{88}\Pi_{88})-\mathcal{K}_{08}\Pi_{08}(2-\mathcal{K}_{08}\Pi_{08})-(\mathcal{K}_{08})^2\Pi_{00}\Pi_{08}-\mathcal{K}_{00} \mathcal{K}_{88} (\Pi_{08})^2\,, 
		\end{equation} 
	\end{widetext}
	where we have hidden the dependency of $p$ on the right-hand side to increase overview and ignored its matrix structure. In FIG.~\ref{fig:eta_prop_} the real and imaginary part of the inverse propagator is shown as a function of the meson momentum $(p_0,\vec p=0)$. Note that the imaginary part of the inverse propagator only comes from the related polarisation loops, cf. appendix~\ref{app:polarisation_loops}. In fact, while the roots of the real part are in general related to bound states of the collective excitation, the imaginary part gives rise to decay processes: Since the NJL model does not contain confinement, mesons are allowed to decay into their constituent quarks. Hence, within the propagator a non-vanishing imaginary part shows up at $p_0=2m_u$ and leads to a kink in the real part. In the case of the parameter set used in FIG.~\ref{fig:eta_prop_} this happens right after the $\eta$-root at $600\,\text{MeV}$. Due to the vanishing imaginary part, we will refer to the region within the inverse propagator for which $p_0<2m_u$ holds as the region of stability in this paper. The second interesting region emerges for $p_0\geq2m_s$, when the decay into a strange--anti-strange pair becomes possible.
	
	\begin{figure}[t]
		\centering
		\includegraphics[width=0.48\textwidth]{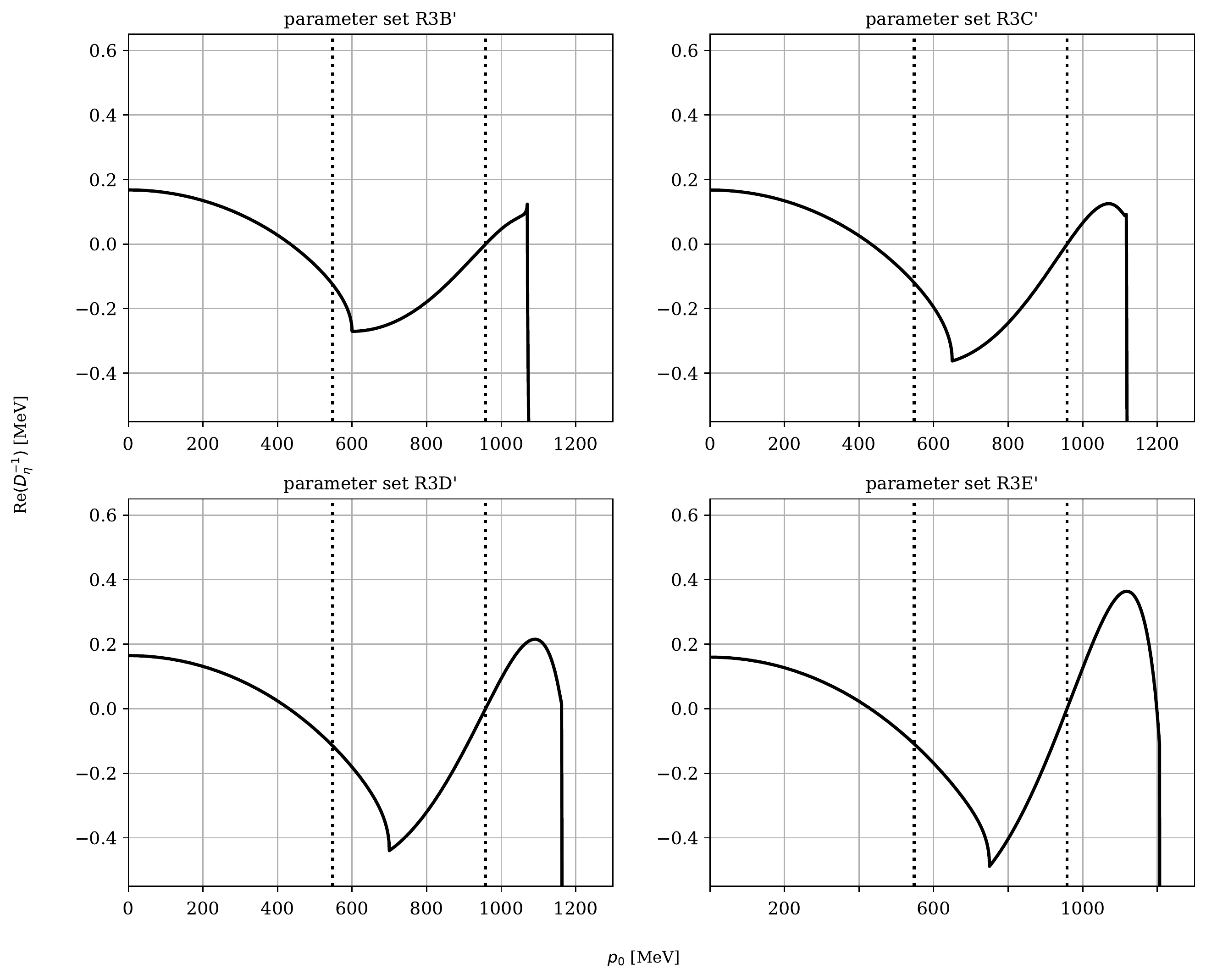}
		\caption{\label{fig:eta_prop_multi} Real part of the inverse $\eta$-propagator as function of $p_0$ for different parameter sets in the Pauli-Villars regularisation method with 3 regulators, cf. TABLE~\ref{tab:parameter_set_PV3R_etaprime}. The vertical lines mark the masses of the $\eta$ and $\eta'$-meson.}
	\end{figure}
		
	In FIG. \ref{fig:eta_prop_multi} the real part of the inverse $\eta$ propagator is shown for three regulators in the PV regularisation scheme. As expected the region of stability within the inverse propagator becomes larger for increasing constituent mass. On the other side we find that constituent up masses lower than $m_{\eta}/2$ lead to an instability of the $\eta$-meson, where it can decay into a pair of up and anti-up quarks. Within the three-flavour NJL model this can be seen as a ``natural” boundary for a reasonable construction of parameter sets. However, we will later come back to this discussion in order to explain our approach to prefer the $\eta'$-meson mass as a fitting parameter instead of the $\eta$-meson mass, cf. section \ref{sec:parameter_sets}.
	
	The previously described interpretation of the real and imaginary part holds for all meson propagators within all regularisation methods. In particular, for the $\eta$ propagator one have to include the imaginary part of the polarisation loops in order to obtain the correct behaviour of the inverse propagator, cf. section \ref{sec:treatment_of_imag}.\\
	
	We want to mention at this point, that without the t'Hooft interaction term no $\eta$-$\eta'$- splitting could be found since the $U_A(1)$-symmetry is not broken. This can be directly seen under virtue of the upper discussion: For $K=0$ the flavour mixing coupling, i.e. equation~\eqref{equ:coupling_eta}, vanishes, while the other couplings reduces to the one in the original four-point interaction Lagrangian. Therefore, the off-diagonal elements of the interaction kernel in~\eqref{equ:eta_matrices} also become zero. Note that the off-diagonal elements of the polarisation loop part do not vanish.
	Taking a closer look into equation~\eqref{equ:det_equation}, the matrix part related to $\eta_0$ and $\eta_8$ is still not fully diagonal. After some algebra the inverse propagator of $\eta$ reduces for $K=0$ to 
	\begin{equation}
		D_{\eta}^{-1}=(1-2G\Pi_{\pi}) (1-2G\Pi_{\pi}^{'})\,,
	\end{equation}
	where we recognize the polarisation loop of the pion $\Pi_{\pi}$. The other polarisation loop $\Pi_{\pi}^{'}$ is equivalent to the pion one, except that it depends on the constituent mass of the strange quark $m_s$ rather than $m_u$. Compared to the inverse pion propagator~\eqref{equ:pion_prop}, it follows from the upper equation, that one root is equal to the pion mass. Therefore, in the case of a not broken $U_A(1)$-symmetry the pion and $\eta$ state are degenerate as expected. The second root can be related to a pion-like state of two strange quarks and has no further meaning.

	\subsection{Pion decay constant}\label{subsec:pdc}
	\begin{figure}[t]
		\centering
		\includegraphics[width=0.28\textwidth]{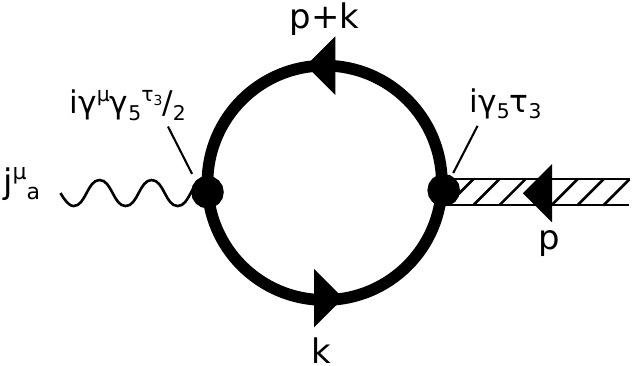}
		\caption{Pion decay Feynman diagram: The pion-decay constant is a result of the coupling between the pion (right-hand side) and the axial current.}
		\label{fig:decayconstant}
	\end{figure}
	The observed weak decay of the charged pions into a muon and a muon-neutrino can be described as the transition probability of a pion decaying into the hadronic vacuum~\cite{burgess2007standard}. Although the electromagnetic decay of $\pi^0$ into two photons is favoured, the weak decay constant will be calculated with the vertex structure of the uncharged pion for simplicity and the fact that we make use of the isospin-limit.\\
	Due to parity conservation, the pion is only allowed to decay through the axial current. It follows that the decay constant $f_{\pi}$, determining the strength of the chiral symmetry breaking \cite{Peskin1995}, can be defined through
	\begin{equation} \label{equ:piondecay}
		\bra{0}A^{\mu}_a(x)\ket{\pi_b(p)}=if_{\pi}p^{\mu}\delta_{ab}e^{ipx}\,,
	\end{equation}
	with the corresponding Feynman diagram shown in FIG.~\ref{fig:decayconstant}. After a carefull evaluation the pion decay constant reads
	\begin{equation} \label{equ:piondecayconstant}
		f_{\pi}=-\left.4N_c \, g_{\pi \bar \psi \psi} \, m_u \, I_2(p^2,m_u)\right|_{p^2=m_{\pi}^2}\,.
	\end{equation}
	where integral $I_2$ defined in appendix~\ref{app:integrals}.

	\section{Regularisation methods}
	\subsection{Sharp momentum cutoff}
	\subsubsection{Four momentum cutoff}
	In the four-momentum cutoff method the euclidean four-momentum integration is restricted to a certain cutoff $\Lambda$. Therefore one has to perform a Wick-rotation in order to rewrite the integrand into euclidean space and transform it to spherical coordinates for a four-dimensional sphere. The result can then be treated with standard techniques, cf. appendix~\ref{app:integrals}.
	\subsubsection{Three momentum cutoff}
	Compared to a four-momentum cutoff, the a three-momentum cutoff does not restrict the integration domain of the four-momentum space completely rather than restrict the spatial component to a cutoff $\Lambda$. Hence, the time component can be evaluated with the residue theorem directly. The remaining three dimensional integral then can be transformed into spherical coordinates without any issues. Here, the radial integration is restricted to radii $|\textbf{k}|<\Lambda$, cf. appendix~\ref{app:integrals}. This scheme obviously breaks Lorentz invariance.
	\subsection{Pauli-Villars cutoff} \label{subsec:Pauli-Villars_cutoff}
	Now, we want to introduce a method to handle divergent integrals described by Pauli and Villars in 1949 and give an overview of its application. Since we just wanted to give short introduction, a more detailed discussion can be found in references \cite{pauli1949invariant, njl-klevansky, njl-buballa}.
	
	While the previous introduced methods use a sharp cutoff to handle the divergent integrals, PV regularisation is related to the proper time regularisation and adds additional terms that behave like the integrand itself for large momenta, but only have a small contribution for small momenta~\cite{moeller-master}. Therefore, the number of additional terms depends on the degree of divergence of the considered integral. Compared to the earlier introduced three-momentum cutoff, this method preserves the Lorentz invariance. Another advantage of PV regularisation compared to hard-cutoff methods is the possibility to make any substitutions within the integrand without taking care of the domain.\\
	Overall, one replaces the original integrand by a weighted sum over new masses, i.e.
	\begin{equation}
		\fourint f(M,k)\longrightarrow \fourint \sum_{j=0}^N c_j f(M_j,k) \label{equ:PVR1}\,,
	\end{equation}
	where $N$ determines the highest order of divergence.
	In general one chooses the coefficient $c_0=1$ and $M_0=M$ in order to get the original version of the integral. The modified masses $M_j(M, \Lambda)$ are given by 
	\begin{equation}
		M_j^2=M^2+j\cdot \Lambda^2\,.\label{equ:PVR-M}
	\end{equation}
	This choice is not unique, and one can of course make a different ansatz. Here, the (soft) cutoff $\Lambda$ enters the model as a new free parameter.
	
	To ensure that all divergences are treated, the coefficients $c_j$ for the additional terms in equation~\eqref{equ:PVR1} have to fulfil a set of equations given by~\cite{Ossola:2003ku}
	\begin{equation} \label{equ:PV_conditions}
			\sum_{j=0}^{N} c_j \left(M_j^2\right)^n = 0\,\text{ for }\, n=(0,1,\dots, N-1)\,.
	\end{equation}	
	The set~\eqref{equ:PV_conditions} determines the coefficients completely. In appendix~\ref{app:integrals} an example for a Pauli-Villars regularized integral is displayed.
	
	In our case, where the highest order of divergence is quadratic, we need two additional regulators, i.e. $N=2$, within the upper condition. This leads to two conditional equations for the coefficients of the PV method. However, in thermodynamic discussions of QCD within the NJL model framework three counter terms, i.e. $N=4$, are needed typically to control the divergences of the thermodynamic potential. Then, an additional equation for $n=2$ enters the set of equations for the $c_j$. The resulting coefficients are shown in~TABLE~\ref{tab:PV_coef}.
	Note that we will treat the three regulator case in the following discussions analogously to the two regulator one.
	
	\begin{table}[t]
		\caption{\label{tab:PV_coef} Coefficients of the PV regularisation method for two and three regulators.}
		\begin{ruledtabular}
			\begin{tabular}{c|cccc}
				\# regulator terms&$c_0$&$c_1$&$c_2$&$c_3$\\
				\hline
				2&1&-2&1&\\
				3&1&-3&3&-1\\
			\end{tabular}
		\end{ruledtabular}
	\end{table}

	\section{Parameter sets for different regularisation methods}
	\label{sec:parameter_sets}
	\begin{table}[t]
		\caption{\label{tab:phys_val_for_params} Values for the physical observables on which the parameters are fixed.}
		\begin{ruledtabular}
			\begin{tabular}{lllll}
				$m_{\pi}$ [MeV]&$m_K$ [MeV]&$m_{\eta}$ [MeV]&$m_{\eta'}$ [MeV]&$f_{\pi}$ [MeV]\\
				\hline
				135.&498.&548.&958.&92.2\\
			\end{tabular}
		\end{ruledtabular}
	\end{table}
	In the three flavour NJL framework we have to fix the parameters of the model to certain physical observables. In general, the parameters are the coupling $G$ and $K$ of the four-point and six-point interaction, respectively, the cutoff $\Lambda$ and the three current masses of the up-, down- and strange-quark as well. In our approach where up- and down-quarks are degenerate, we only have five free parameters.
	
	\begin{table}[t]
		\caption{\label{tab:perfect_solution} Parameter sets for all regularisation schemes fixed to all physical observables. They are named after the regularisation schemes: C4~- four momentum cutoff, C3~- three momentum cutoff, R2~- PV with two regulators, R3~- PV with three regulators}
		\begin{ruledtabular}
			\begin{tabular}{ll|llll}
				&&[C4i]&[C4ii]&[C3i]&[C3ii]\\
				\hline &&&&&\\
				$m_u$ & $[\text{MeV}]$ & 263.043 & 275.65 & 261.941 & 277.028\\
				$m_s$ & $[\text{MeV}]$ & 399.55 & 412.222 & 368.908 & 384.93 \\
				$\Lambda$  & $[\text{MeV}]$ & 898.011 & 863.096 & 736.846 & 701.415 \\
				$G\Lambda^2$ &  & 0.926 & 0.622 & 0.634 & 0.582 \\
				$K\Lambda^5$ &  & 292.407 & 323.762 & 44.938 & 45.93 \\
				\hline &&&&&\\
				$m_{0,u}$ & $[\text{MeV}]$ & 6.144 & 6.548 & 4.3 & 4.607 \\
				$m_{0,s}$ & $[\text{MeV}]$ & 167.953 & 175.442 & 125.147 & 130.728 \\
				\hline \hline &&&&&\\[-0.25cm]	
				&&[R2i]&[R2ii]&[R3i]&[R3ii]\\
				\hline &&&&&\\
				$m_u$ & $[\text{MeV}]$ & 262.694 & 275.855 & 268.178 & 274.201 \\
				$m_s$ & $[\text{MeV}]$ & 397.595 & 411.034 & 421.644 & 426.545 \\
				$\Lambda$  & $[\text{MeV}]$ & 770.037 & 738.57 & 879.603 & 863.981 \\
				$G\Lambda^2$ &  & 0.706 & 0.5 & -0.064 & -0.593 \\
				$K\Lambda^5$ &  & 126.45 & 139.422 & 589.039 & 657.881 \\
				\hline &&&&&\\
				$m_{0,u}$ & $[\text{MeV}]$ & 6.004 & 6.418 & 7.72 & 7.97 \\
				$m_{0,s}$ & $[\text{MeV}]$ & 164.824 & 172.497 & 202.777 & 207.3 \\			
			\end{tabular}
		\end{ruledtabular}
	\end{table}		
	
	The parameters are fitted to pion mass, kaon mass, $\eta$ and $\eta'$ masses 
	\begin{align}
		&\text{Re }D_{\pi}^{-1}(m_{\pi}^2) = \text{Re }D_{K}^{-1}(m_{K}^2)\stackrel{!}{=}0 \label{equ:inverse_prop_pion}\\
		&\text{Re }D_{\eta}^{-1}(m_{\eta}^2) = \text{Re }D_{\eta}^{-1}(m_{\eta^{\prime}}^2) \stackrel{!}{=} 0\label{equ:inverse_prop_eta}
	\end{align}
	and pion decay constant~\eqref{equ:piondecayconstant}. If not stated otherwise, the full complex character of the emerging integrals and therefore the inverse propagator will be used. The values of these physical observables are shown in TABLE~\ref{tab:phys_val_for_params}. Note that within the related equations the constituent quark mass needs to be applied. Therefore, the two self-consistent gap-equations, cf. equation~\eqref{equ:gapequ}, have to be solved simultaneously as well. Of course, one can imagine other approaches where different observables can be fixed to create a set of parameters.	
	
	As a matter of fact, it is possible to find two such ``perfect" parameter sets for each regularisation method, cf. TABLE~\ref{tab:perfect_solution}. The corresponding values of $m_u$ are comparatively small. One parameter-set of each scheme is actually related to an unstable $\eta$-meson since $2m_u<m_{\eta}$. In the case of Pauli-Villars regularisation with three regulators the four-point couplings are even negative. Hence, these special parameter-sets will not always be appropriate which is why we decided to create various ones by varying one parameter. In our case we use the constituent mass of the up-quark $m_u$ as a free parameter between $275\,\text{MeV}$ and $375\,\text{MeV}$.
	
	First, we decided to fit the remaining four parameters to pion mass, kaon mass, $\eta$ mass and pion decay constant. The calculated $\eta'$ masses, besides the ``perfect" solutions around $m_u = m_{\eta}/2$, lie well above the experimental value. Using only the real part of integrals leads to even higher masses, cf. section~\ref{sec:treatment_of_imag}. Since the corresponding roots of the inverse $\eta$-propagator lie in every case in the unstable region, cf. section~\ref{subsec:meson_prop}, we decided not to trust the calculated $\eta'$ masses. It seems to be more convenient for all regularisation methods to use it as a fixed point for the parameters. Then, the calculated mass of the $\eta$-meson always lies within the region of stability. For the different regularisation methods in case of fixed $\eta'$ masses, the corresponding mass of the $\eta$-meson is given. For up-quark constituent masses between $275\,\text{MeV}$ and $375\,\text{MeV}$ the masses of the $\eta$-meson is relatively close to its experimental value.
	
	Our results for different regularisation methods are presented in section~\ref{subsec:sharp_cutoff_parameter} and~\ref{subsec:PV_parameter}. All of them have been calculated numerically using \textit{Mathematica 12.0}~\cite{Mathematica}. The determination of parameter-sets has been done in two steps: First, we specified the value of $m_u$ and fixed the remaining parameters, i.e. $m_s$, $\Lambda$, $G$, and $K$, to $m_{\pi}$, $m_K$, $f_{\pi}$ and either $m_{\eta}$ or $m_{\eta'}$. Second, with the parameters at hand, we computed $m_{u,0}$, $m_{s,0}$ and either $m_{\eta'}$ or $m_{\eta}$. The couplings $G$ and $K$ are multiplied by appropriate powers of $\Lambda$ to obtain dimensionless numbers. 

	\subsection{Sharp momentum cutoff}
	\label{subsec:sharp_cutoff_parameter}
	\subsubsection{Four momentum cutoff}
	\label{subsubsec:four_momentum_cutoff}
	\begin{figure}[t]
		\centering
		\includegraphics[width=0.48\textwidth]{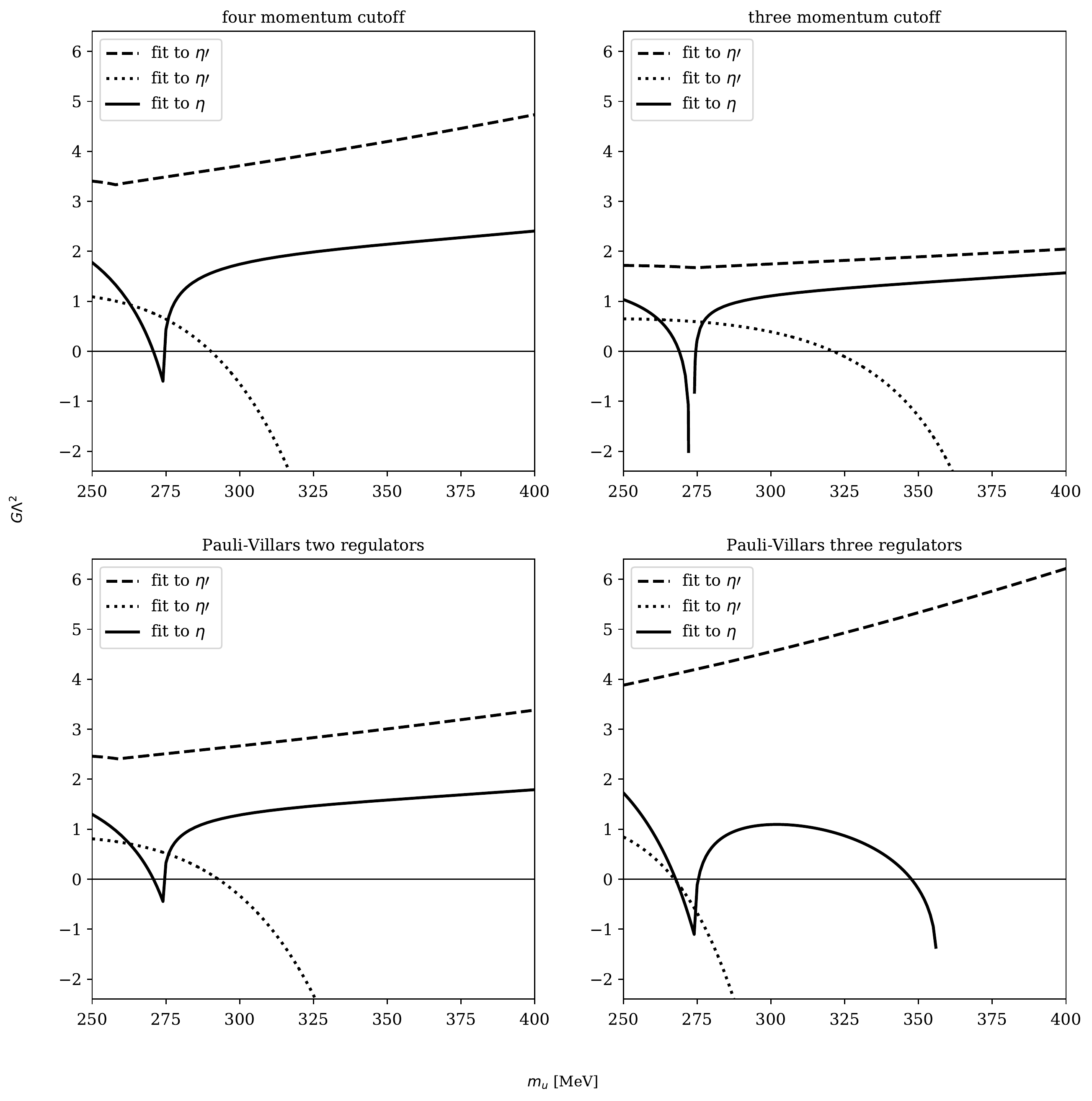}
		\caption{\label{fig:G_as_fct_of_mu} $G\Lambda^2$ as function of $m_u$ for different regularisation methods, fit to $\eta$ (solid), $\eta^{\prime}$ (dashed) and $\eta^{\prime}$ \textit{alternative solution} (dotted).}
	\end{figure}
	\begin{figure}[t]
		\centering
		\includegraphics[width=0.48\textwidth]{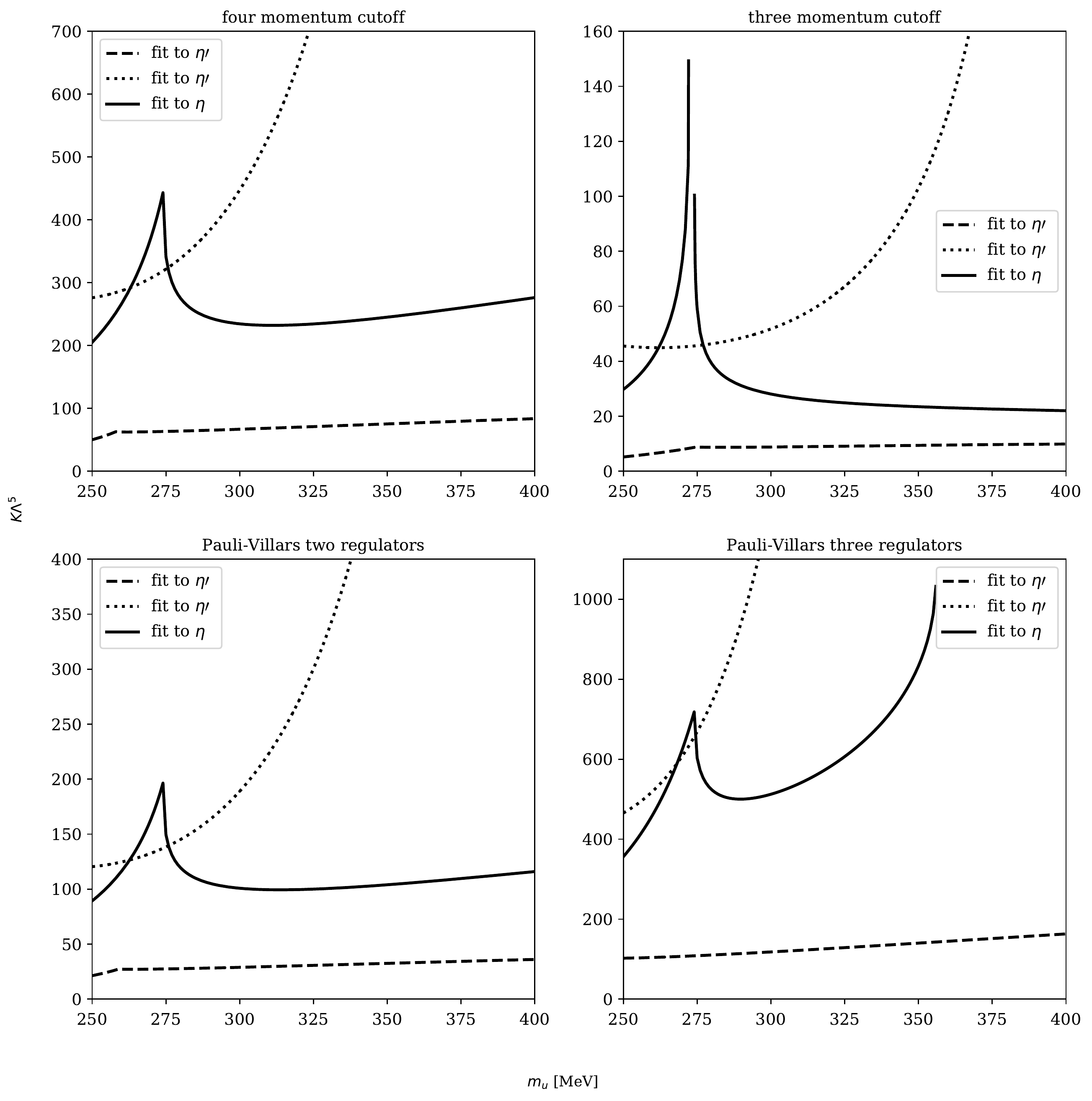}
		\caption{\label{fig:K_as_fct_of_mu} $K\Lambda^5$ as function of $m_u$ for different regularisation methods, fit to $\eta$ (solid), $\eta^{\prime}$ (dashed) and $\eta^{\prime}$ \textit{alternative solution} (dotted). Note that the vertical axes have different scales.}
	\end{figure}
	\begin{figure}[t]
		\centering
		\includegraphics[width=0.48\textwidth]{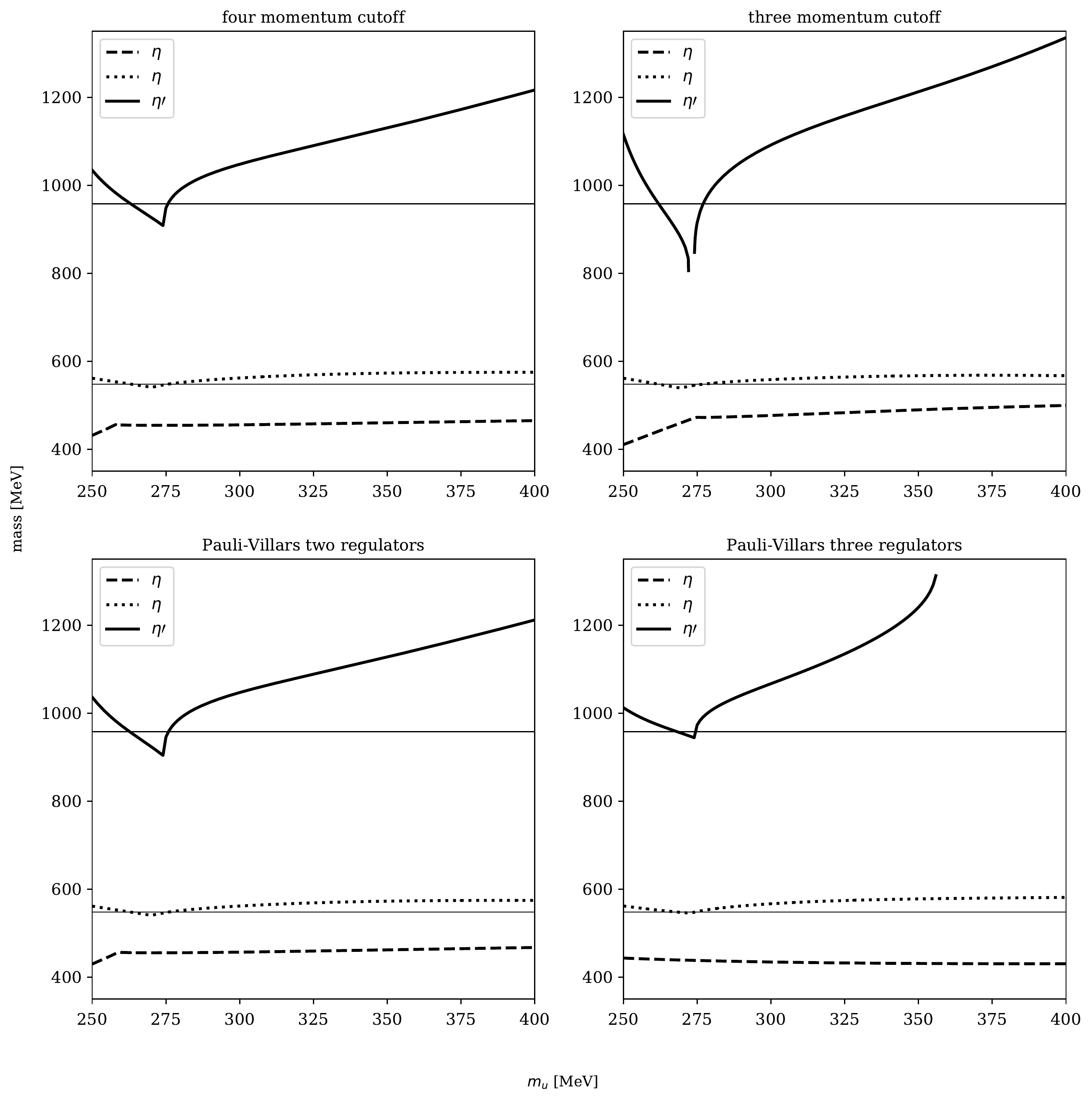}
		\caption{\label{fig:Meta_as_fct_of_mu} $m_{\eta}$ or $m_{\eta^{\prime}}$ as function of $m_u$ for different regularisation methods, fit to $\eta$ (solid), $\eta^{\prime}$ (dashed) and $\eta^{\prime}$ \textit{alternative solution} (dotted). Displayed is the mass \textit{not} fitted to. Horizontal lines show the experimental values for $m_{\eta}$ and $m_{\eta^{\prime}}$.}
	\end{figure}
	Using the four-momentum cutoff scheme, we find for given $m_u$ one solution when fixing the $\eta$ and two solutions when fixing the $\eta'$ mass. The corresponding parameter sets are displayed in TABLE~\ref{tab:parameter_set_4C_eta} and in TABLE~\ref{tab:parameter_set_4C_etaprime} and~\ref{tab:parameter_set_4C_etaprime_II}, respectively. We decided not to display the calculated values for $m_{\eta'}$ for fixed $\eta$ mass because it lies in the unstable region of the propagator, cf. section~\ref{subsec:meson_prop}. \\
	The dependence of the couplings $G$ and $K$ on the constituent mass of the up-quark is shown in FIG.~\ref{fig:G_as_fct_of_mu} and~\ref{fig:K_as_fct_of_mu} in the upper left panels, respectively. Intersections of solutions correspond to ``perfect" solutions for all physical observables, cf. parameter sets [C4i] and [C4ii] in \mbox{TABLE~\ref{tab:perfect_solution}.} \\
	Having a closer look at the solution for fixed $\eta$ mass, we find that the couplings show only marginal dependence on $m_u$ except for the domain around \mbox{$m_u = m_{\eta}/2 = 274\,\text{MeV}$.} For values $m_u < 274\,\text{MeV}$ the $\eta$-meson can decay into a quark-antiquark pair resulting in a kink in the real part of the meson propagator. This kink leads to the interesting behaviour of the couplings where, of course, it also emerges: While the six-point coupling just grows to large values in this domain the four-point one becomes negative, i.e. repulsive. Afterwards, the values of both couplings stabilize again. Therefore, we consider only parameter sets with $m_u \geq 275\,\text{MeV}$. \\
	In the case of fixed $\eta'$ mass the situation is different: One solution shows only marginal dependence on $m_u$ over a wide range while the other one shows a strong dependence. In the following, the second solution will be referred to as \textit{alternative solution}. For $m_u>290\,\text{MeV}$ the four-point coupling of the \textit{alternative solution} becomes negative and tends fast towards highly negative values with increasing constituent mass of the up-quark. Of course, in order to compensate that the six-point coupling have to increase fast. This behaviour may be a sign for the solution being unphysical but further analysis is required here. It is noteworthy that the ``perfect" solutions belong to the \textit{alternative solution} and, therefore, have to treated with some care as well.\\
	In FIG.~\ref{fig:Meta_as_fct_of_mu} the mass of the $\eta$- and $\eta'$-meson depending on which one has been used for the parameter fit is shown as a function of $m_u$. Here, we also see that only the \textit{alternative solution} in the case of a fit to the $\eta'$-mass will lead to a ``perfect'' parameter set, where both physical $\eta$-masses are reproduced. Meanwhile the other $\eta'$-mass fit underestimates the mass of the $\eta$-meson constantly and does not cross the line of the physical value of $m_{\eta}$. However, both $\eta$-masses based on fits to $\eta'$ do not have a strong dependence of the chosen value of $m_u$. Since this observation holds for all used regularisation methods it already suggests that the $\eta'$-mesons should be used in order to fit parameters due to a broader range of its mass.
	
	The authors grade the standard parameter set fixed to the $\eta'$ mass as the most promising one. All couplings stay in reasonable intervals for a wide range of the constituent mass of the up-quark and show no kinks or discontinuities. Nevertheless, the calculated $\eta$ masses are roughly $100\,\text{MeV}$ below the experimental value.

	\subsubsection{Three momentum cutoff}
	\label{subsubsec:three_momentum_cutoff}
	
	Using the three-momentum cutoff scheme, we find results similar to the four-momentum cutoff ones, i.e. one solution when fixing the $\eta$, cf. TABLE~\ref{tab:parameter_set_3C_eta}, and two solutions when fixing the $\eta'$ mass, cf. TABLE~\ref{tab:parameter_set_3C_etaprime} and~\ref{tab:parameter_set_3C_etaprime_II}. \\
	The dependence of the couplings $G$ and $K$ on the constituent mass of the up-quark is shown in FIG.~\ref{fig:G_as_fct_of_mu} and~\ref{fig:K_as_fct_of_mu} in the upper right panels, respectively. Intersections of solutions correspond to ``perfect" solutions for all physical observables again, cf. parameter sets [C3i] and [C3ii] in TABLE~\ref{tab:perfect_solution}. \\
	Having a closer look at the solution for fixed $\eta$ mass, we again find that the couplings show only marginal dependence on $m_u$ except for the domain around $m_u = m_{\eta}/2$. Here, we observe an unique problem: For values close to $m_u = 274\,\text{MeV}$, we were not able to find any solution when fixing the $\eta$ mass. The authors believe that this is just a numerical problem and expect a similar kink as in the four-momentum cutoff case. \\
	In the case of fixed $\eta'$ mass the situation is different again: One solution shows only marginal dependence on $m_u$ over a wide range while the other one shows a strong dependence. In the following the second solution is referred to as \textit{alternative solution}. For $m_u>321\,\text{MeV}$ the four-point coupling of the \textit{alternative solution} becomes negative and tends fast towards highly negative values with increasing constituent mass of the up-quark. Of course, in order to compensate that the six-point coupling have to increase fast. This behaviour is similar to the four-momentum cutoff case and implies the same conclusions.
	
	Here, the most promising parameter set seems to be the standard one fixed to the $\eta'$ mass again. Once more, the calculated $\eta$ masses are roughly $80\,\text{MeV}$ below the experimental value.
	
	Comparing the four and three-momentum cutoff schemes, we find that the four-point coupling $G$ is roughly twice as big in the four-momentum case over the whole range of the up-quark constituent mass. For the \textit{alternative solutions} this statement holds only for small $m_u$. Making the same statement for the six-point coupling $K$ is more complicated, the factor is of order $10$ but varies with $m_u$. Besides the discontinuity in case of fixed $\eta$ mass both schemes lead to similar parameter sets, which can also be seen in FIG.~\ref{fig:Meta_as_fct_of_mu}. The four-momentum cutoff scheme provides somewhat larger current quark masses.
	
	\subsection{Pauli-Villars cutoff}
	\label{subsec:PV_parameter}
	\subsubsection{Two regulators}
	\label{subsubsec:two_regulators}	
	
	Analogously to the previous discussed sharp cutoff methods we have calculated parameter sets by fixing the $\eta$ and $\eta'$ mass, respectively with the Pauli-Villars regularisation method with two regulators. These parameter sets are shown in TABLE~\ref{tab:parameter_set_PV2R_eta} to~\ref{tab:parameter_set_PV2R_etaprime_II} again, by varying $m_u$.\\
	Taking a look at FIG.~\ref{fig:G_as_fct_of_mu}, FIG.~\ref{fig:K_as_fct_of_mu} and FIG.~\ref{fig:Meta_as_fct_of_mu} (left lower panels) where the dependence of the coupling constants $G$, $K$ and $m_{\eta}$ (respectivley $m_{\eta'}$) to the mass of the up-quark are shown, one notes that the behaviour is qualitatively nearly identical to the one for the four-momentum cutoff one. The only difference lies in the fact that the coupling of the six-point term, i.e. $K$, for the Pauli-Villars methods is roughly half as high as the ones we find for the four-momentum cutoff scheme. The four-point couplings are nearly the same for both methods. Therefore we want to refer to section \ref{subsubsec:four_momentum_cutoff}. Intersections correspond to ``perfect" solutions again, cf. parameter sets [R2i] and [R2ii] in TABLE~\ref{tab:perfect_solution}.\\
	However, it seems that the Pauli-Villars regularisation needs a stronger coupling in the six-point term, in order to ensure the $\eta$-$\eta'$-splitting within the given restrictions of the up-quark mass. This may be related to the fact that in the Pauli-Villars regularisation higher momenta of the quark-loops are allowed and not ignored completely because of a sharp cutoff. Besides the different ansatz of both methods, the four-momentum cutoff and the Pauli-Villlars scheme even lead to quantitative similar parameter sets, cf. TABLE~\ref{tab:parameter_set_4C_eta} to \ref{tab:parameter_set_3C_etaprime_II} and TABLE~\ref{tab:parameter_set_PV2R_eta} to \ref{tab:parameter_set_PV2R_etaprime_II}: Here, the $\eta$ mass is around \mbox{$100\,\text{MeV}$} lower than experimentally expected as discussed earlier.\\
	
	Again, the \textit{alternative solution} represents a set of parameters where the $\eta$ mass is in good agreement with the experimental value but corresponds to negative values of the four-point coupling and unreasonable high values for $K$ at some certain value for $m_u$.
	
	\subsubsection{Three regulators}
	\label{subsubsec:three_regulators}
	
	Beside the Pauli-Villars regularisation scheme with two additional regulator terms, we have also investigated this method with three regulators. The latter one is often used in the NJL model for in-medium discussions and hence should be taken into account as well. \\
	The parameters sets, which have been calculated analogously to the previous ones, are shown in TABLE~\ref{tab:parameter_set_PV3R_eta} to \ref{tab:parameter_set_PV2R_etaprime_II}. Again, we fixed the parameters to the mass of the $\eta$-meson and $\eta'$-meson, respectively. In order to compare the results to the other methods the lower right panel of FIG.~\ref{fig:G_as_fct_of_mu} and~\ref{fig:K_as_fct_of_mu} shows the behaviour of the four-point and six-point couplings in dependence of the up-quark mass. Intersections correspond to ``perfect" solutions, cf. parameter sets [R3i] and [R3ii] in TABLE~\ref{tab:perfect_solution}.\\
	One immediately notice the additional drop at around $350\,\text{MeV}$ in the case where we fitted to the $\eta$ mass. It turns out that we do not find reasonable parameter sets for $m_u > 355\,\text{MeV}$. Therefore, in FIG.~\ref{fig:Meta_as_fct_of_mu}, no $\eta'$-mass can be calculated. Nevertheless, the fact that equal values of the couplings $G$ and $K$ do not have a critical impact on the related $\eta'$-mass before this point is strongly noticeable. It seems that compared to the other regularisation methods, the Pauli-Villlars scheme with three regulators has the strongest dependence on the quark masses in terms of the $\eta$-meson. We assume this behaviour is related to the third PV regulator but further studies are required to confirm this conjecture. \\
	The bare values we find for $K$ are in this case way bigger compared to all other regularisation schemes.
	
	\section{Treatment of imaginary parts}
	\label{sec:treatment_of_imag}
	
	\begin{figure}[t]
		\centering
		\includegraphics[width=0.48\textwidth]{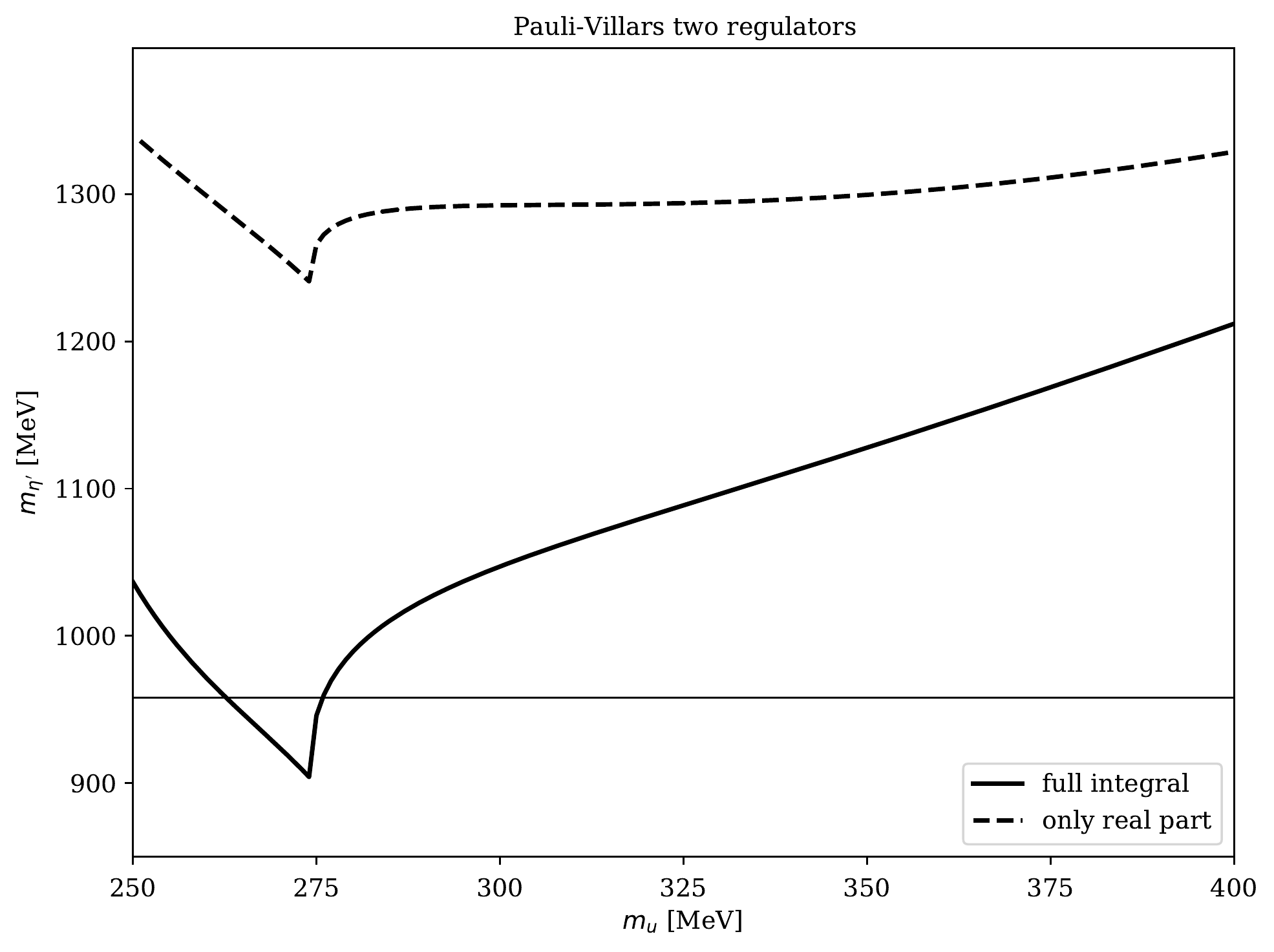}
		\caption{\label{fig:PV2R_real_part_vs_full_integral_metaprime} $m_{\eta^{\prime}}$ as function of $m_u$ in the Pauli-Villars regularisation method with 2 regulators fit to $\eta$ mass, using full integrals (solid) and only real part (dashed).}
	\end{figure}
	
	As already mentioned earlier, the imaginary part of the polarisation loops plays an important role for the inverse propagators of the mesons.\\
	In the literature we found that some authors, cf. reference~\cite{HATSUDA1994221,Rehberg:1995kh,Kohyama:2016fif} used the real part of the integrals and neglected the imaginary part or just fitted the $\eta'$ mass to high enough values of $m_u$. While in the two flavour-case this shouldn't be a problem because the emerging mesons, i.e. pions, only have small masses, this will not hold for the three flavour case anymore.\\
	In section~\ref{subsec:meson_prop} we have seen that the $\eta$ propagator of course is sensitive to a non-vanishing imaginary part of the integrals.\\

	In TABLE~\ref{tab:full_vs_real_integral_PV2R_eta_etaprime_masses} the parameter sets for various fixed $m_u$ and $m_{\eta'}$ are shown. Choosing the Pauli-Villars scheme with two regulators to be representative for all methods, cf. FIG.~\ref{fig:PV2R_real_part_vs_full_integral_metaprime}, it can be directly seen that the corresponding mass for the $\eta'$-meson is way to high if we only take the real part for the calculations. Therefore, no ``perfect'' solution can be found when using only the real part. This comes from the fact that the inverse $\eta$ propagator contains multiple products of complex numbers, i.e. kinds of polarisation loops (cf. equation~\eqref{equ:eta_prop}). Hence, the imaginary part of a certain polarisation loop indeed contributes to the real part of the hole propagator. Since the root of the inverse $\eta$-propagator related to $\eta'$ lies in the vicinity of the strange--anti-strange decay window, we have recalculated the parameter sets for the Pauli-Villars method with two regulators for fixed values of $m_s$. In order to increase the understanding TABLE~\ref{tab:full_vs_real_integral_PV2R_etaprime} in appendix~\ref{app:real_vs_full_integral} shows different parameter-sets for the two regulator Pauli-Villars regularisation method, where $m_s$ has been fixed instead of $m_u$. Due to the allowed decay into a strange-anti-strange quark pair for values of $m_s < m_{\eta'}/2$, the parameter sets do depend whether the full or only the real part of the integral has been taken into account. However, the decay into two up-quarks is in all of these sets allowed and therefore does not change the parameter sets at all.\\
	Despite the simplicity of this discussion, one has to keep the upper observation in mind when creating or using parameter sets for the three flavour NJL model.
	\begin{table}[t]
		\caption{\label{tab:full_vs_real_integral_PV2R_eta_etaprime_masses} Comparing calculated results for $m_{\eta^{\prime}}$ using full and only real part of integrals (Pauli-Villars scheme with two regulators fit to $\eta$ mass).}
		\begin{ruledtabular}
			\begin{tabular}{l|lllll}
				$m_u \; [\text{MeV}]$ & 275. & 300. & 325. & 350. & 375. \\
				\hline &&&&&\\
				\textit{full integral} &&&&&\\[0.1cm]
				$m_{\eta^{\prime}} \; [\text{MeV}]$ & 945.78 & 1047.06 & 1088.62 & 1127.75 & 1168.78 \\
				&&&&&\\
				\textit{only real part} &&&&&\\[0.1cm]
				$m_{\eta^{\prime}} \; [\text{MeV}]$ & 1265.73 & 1292.40 & 1293.84 & 1299.50 & 1311.21 \\
			\end{tabular}
		\end{ruledtabular}
	\end{table}
	
	\clearpage
	\section{Conclusions and Outlook}		
	All in all, we found for all regularisation methods three different sets of parameters where the mass of the up-quark has been treated as a free parameter.\\
	The different parameters sets for various regularisation methods, are displayed in appendix~\ref{app:parameter_sets}. Here, one parameter set is fitted to the $\eta$ mass and two to the $\eta'$ mass. The ones fitted to $\eta$ show some interesting behaviour in terms of the couplings: While the fits to the $\eta'$ mass stay smooth, the ones fitted to $\eta$ show a dip around $m_u = m_{\eta}/2 = 274\,\text{MeV}$. This is related to the allowed decay into a quark-antiquark pair which leads to a non-vanishing imaginary part of the meson propagator. Hence, the analytical behaviour of the propagator is different before and after this region.\\
	It seems that the parameter fit to the $\eta'$-meson is more reasonable although its mass always lies outside the region of stability. Therefore, the related meson propagator always has a non-vanishing imaginary part. Nevertheless, under virtue of FIG.~\ref{fig:G_as_fct_of_mu} and~\ref{fig:K_as_fct_of_mu} the parameter sets coming from this ansatz are steady in terms of their changes over a wide range of given $m_u$. In addition, no negative values for $G$ have been found within the considered range of $m_u$. However, the parameter-sets of the so-called \textit{alternative solutions}, which have been also fitted to $\eta'$, lead to negative values of the four-point coupling $G$ at a certain point.\\	
	Besides, we found that the Pauli-Villars regularisation method with three regulators has to be treated with some care when fitted to the $\eta$ mass. For $m_u > 355\,\text{MeV}$ it was not possible to find reasonable parameter. Here, further studies are required. \\
	We have already mentioned the in-medium discussions of the NJL model and the corresponding thermodynamic potential, cf. section \ref{subsubsec:three_regulators} . In fact, this ansatz can also be used to calculate parameter sets for the NJL model: Hereby one calculates the thermodynamic potential of the model via its partition function. The (physical) ground state of the model is then given by the global minimum of the thermodynamic potential. Of course one can do this in the limit of vanishing temperature and chemical potential, i.e. in the vacuum as well. Then the results can be compared with the ones of this work directly.\\		
	As an example, a short look at this method, already showed that the corresponding global minimum of the thermodynamic potential for the \textit{alternative solution} is not as deep as the one for the other parameter sets. This holds for all regularisation methods. However, we want to emphasise at this point that this may not have any importance for a certain study within the NJL model framework, but can be seen as an indicator that the \textit{alternative solution} have to be treated with some care.\\
	
	The ``perfect" parameter sets however, which reproduces the $\eta$ and $\eta'$ mass correctly, correspond to intersections of the fits to each mass separately, cf. TABLE~\ref{tab:perfect_solution} and FIG.~\ref{fig:G_as_fct_of_mu} and~\ref{fig:K_as_fct_of_mu}. Here, we have found that only intersections of the \textit{alternative solution} and the fit to $\eta$ lead to results. Hence, two possible sets can be identified: Intersections before the dip in the \textit{alternative solutions} corresponds to a set with a stable $\eta$-meson. The other intersections meanwhile corresponds to a set with an unstable $\eta$-meson.\\
	
	Within the upper discussion we have investigated the impact of neglecting the imaginary part of the integrals and discussed it exemplary on the Pauli-Villars method with two regulators, cf. section~\ref{sec:treatment_of_imag}. When fitting to the $\eta'$ mass, for $m_s\geq m_{\eta'}/2$ the results do not depend whether one uses the full integral or only the real part. However, smaller $m_s$ will lead to different results, cf. TABLE~\ref{tab:full_vs_real_integral_PV2R_etaprime}. In terms of $m_u$ this means that only for $m_u\geq259$~MeV the obtained parameter sets do not depend on the treatment of the possible imaginary part of the integrals.\\
	When fitting to the $\eta$ mass, neglecting the contribution of the imaginary part in the polarisation loop yields unreasonable high masses for the $\eta'$ meson, cf. TABLE~\ref{tab:full_vs_real_integral_PV2R_eta_etaprime_masses}. In order to choose a wider range of constituent masses and to be sure to have consistent results the fully complex character of the polarisation loops should be taken into account.\\

	Since the NJL model is often used as an effective model of QCD for in-medium studies, a comprehensive discussion of the different regularisation methods could be of particular interest. Especially a comparison of the obtained phase structure for each method and, moreover, with other approaches like FRG or lattice calculations, could give more insights in where the model boundaries lie. In addition other fit parameters like the mixing angle of the $\eta$ and $\eta'$-meson or the kaon decay constant can be used to fit the free parameters of the model. Alternatively one can also used the parameter sets of this work in order to calculate certain physical observables to verify their validity in the context of the NJL model as an effective model of QCD.\\
	
	\begin{acknowledgments}	
		The authors would like to thank Michael~Buballa, Dirk~H.~Rischke and his group of the \textit{Institute for Theoretical Physics} at the \textit{Goethe Universit\"at Frankfurt am Main} for the constructive discussion and helpful comments.\\
		D.K. acknowledges the support of the \textit{Deutsche Forschungsgemeinschaft} (DFG, German Research Foundation) through the collaborative research center trans-regio CRC-TR 211 `Strong-interaction matter under extreme conditions'– project	number 315477589 – TRR 211. The research of D.K. is supported by the \textit{Helmholtz Graduate School for Hadron and Ion Research}. \\
		The research of M.S. is supported by the Cluster of	Excellence \textit{Precision Physics, Fundamental Interactions and Structure of Matter} (PRISMA$^+$, EXC 2118/1) within the German Excellence Strategy (project ID 39083149). 
	\end{acknowledgments}

	\appendix
	
	\section{Integrals} \label{app:integrals}
	
	Evaluating quark condensate \eqref{equ:quark_condensate} and polarisation loops \eqref{equ:polarisation_loops} will lead to the following three integrals
	
	\begin{widetext}
		\begin{align}
			\label{equ:integral_I1}
			I_1(M) &\equiv  i \fourint \frac{1}{k^2 - M^2 + i \epsilon} \, ,
			\\
			\label{equ:integral_I2}
			I_2(p^2, M) &\equiv i \fourint \frac{1}{\left[(p+k)^2-M^2+i\epsilon\right] \left[k^2-M^2+i\epsilon\right]} \, ,
			\\
			\label{equ:integral_I3}
			I_3(p^2, M_1, M_2) &\equiv i \fourint 	\frac{1}{\left[(p+k)^2-M_1^2+i\epsilon\right] \left[k^2-M_2^2+i\epsilon\right]} \, .
		\end{align}
	\end{widetext}
	
	Of course, these integrals are divergent and have to be regularized. E.g. for integral $I_1$, using the four-momentum cutoff scheme and Wick-rotating the integral, we can evaluate it directly and find
	\begin{equation}
		I_1^{C4}(M,\Lambda) = \frac{1}{16 \pi^2} \left(\Lambda^2 + M^2 \ln \left(\frac{M^2}{M^2 + \Lambda^2}\right)\right) .
	\end{equation}
	For the other regularisation methods we perform the $k_0$ integration using Cauchy's residue theorem. The remaining three-dimensional spherical integral can be computed directly. In the three-momentum cutoff case it reads
	\begin{widetext}
		\begin{equation}
			I_1^{C3}(M,\Lambda) = \frac{1}{8 \pi^2} \left(\Lambda \sqrt{M^2 + \Lambda^2} + M^2 \ln \left(\frac{M}{\sqrt{M^2 + \Lambda^2} + \Lambda}\right)\right)
		\end{equation}
	\end{widetext}
	and for Pauli-Villars regularisation scheme
	\begin{equation}
		I_1^{PV}(M,\Lambda) = \frac{1}{16 \pi^2} \sum_{j=0}^N c_j \, M_j^2 \ln(M_j^2) \,,
	\end{equation}
	where $N \in \{2,3\}$ denotes the number of regulators, $c_j$ and $M_j(M,\Lambda)$ are defined in section~\ref{subsec:Pauli-Villars_cutoff}.
	\\ \\
	Integral~\eqref{equ:integral_I3} remains real for $p^2 > (M_1+M_2)^2$ and gets imaginary contributions for $p^2 < (M_1+M_2)^2$. Since we have to evaluate at e.g. $p^2 = m_{\eta}^2$ when fixing the $\eta$ mass, the correspondence between upcoming imaginary part and allowed decay into quark-antiquark pair is obvious.

	\clearpage
	\section{Polarisation loops}\label{app:polarisation_loops}
	
	From equation~\eqref{equ:polarisation_loops} and TABLE~\ref{tab:scat_kernel_pol_loop_3fl} we can directly read of the definition of the pion polarisation loop
	\begin{equation}
		-i\Pi_{\pi}(p^2) \equiv -\fourint\Tr\left[i\gamma_5 \tau_3 S(p+k)i \gamma_5 \tau_3 S(k)\right] \, .
	\end{equation}
	Due to the isospin limit the polarisation loops for all three pions are the same. This justifies the above definition. A straightforward calculation leads to
	\begin{equation} \label{equ:pol_pion_app}
		\Pi_{\pi}(p^2) =  8N_cI_1(m_u)-4N_cp^2I_2(p^2,m_u) \, ,
	\end{equation}
	where we recognize integral $I_1$ and $I_2$ (cf. appendix~\ref{app:integrals}). The kaon polarisation loop is defined as
	\begin{equation}
		-i\Pi_{K}(p^2) \equiv -\fourint\Tr\left[i\gamma_5 \tau_1^+ S(p+k)i \gamma_5 \tau_1^- S(k)\right] \, ,
	\end{equation}
	where $\tau_1^{\pm} \equiv \frac{1}{\sqrt{2}}\left(\tau_1 \pm  i\tau_2\right)$. As in the pion case, all four kaon polarisation loops are degenerated and we find
	\begin{widetext}
		\begin{equation} \label{equ:pol_kaon_app}
			\Pi_{K}(p^2) =  4N_c\left(I_1(m_u)+I_1(m_s)\right)+4N_c\left(\left(m_u-m_s\right)^2-p^2\right)I_3(p^2,m_u,m_s) \, ,
		\end{equation}
	\end{widetext}
	where the more general integral $I_3$ appears instead of $I_2$. As discussed in section~\ref{subsec:meson_properties}, the $\eta$ and $\eta^{\prime}$ are composite particles. Therefore, also off-diagonal polarisation loops are relevant in this channel
	\begin{widetext}
		\begin{align}
			-i\Pi_{00}(p^2) &= -\fourint\Tr\left[i\gamma_5 \tau_0 S(p+k)i \gamma_5 \tau_0 S(k)\right] \, , 
			\\
			-i\Pi_{88}(p^2) &= -\fourint\Tr\left[i\gamma_5 \tau_8 S(p+k)i \gamma_5 \tau_8 S(k)\right] \, ,
			\\
			-i\Pi_{08}(p^2) &= -\fourint\Tr\left[i\gamma_5 \tau_0 S(p+k)i \gamma_5 \tau_8 S(k)\right] \, ,
		\end{align}
	\end{widetext}
	where we dropped the superscript $\text{PS}$ for simplicity. The third polarisation loop is symmetric, i.e. $\Pi_{08} = \Pi_{80}$. An explicit calculation leads to
	\begin{widetext}
		\begin{align} 
			\label{equ:pol_eta_00_app}
			\Pi_{00}(p^2) &=  \frac{8}{3}N_c\left(2 \, I_1(m_u)+I_1(m_s)\right)-\frac{4}{3}N_c p^2 \left(2\,I_2(p^2,m_u)+I_2(p^2,m_s)\right) \, ,
			\\
			\label{equ:pol_eta_88_app}
			\Pi_{88}(p^2) &=  \frac{8}{3}N_c\left(I_1(m_u)+2 \, I_1(m_s)\right)-\frac{4}{3}N_c p^2 \left(I_2(p^2,m_u)+2 \, I_2(p^2,m_s)\right) \, ,
			\\
			\label{equ:pol_eta_08_app}
			\Pi_{08}(p^2) &= \frac{8}{3}\sqrt{2}N_c \left(I_1(m_u)-I_1(m_s)\right)-\frac{4}{3}\sqrt{2}N_c p^2 \left(I_2(p^2,m_u)-I_2(p^2,m_s)\right) \, .
		\end{align}
	\end{widetext}
	All other combinations $(a,b)$ in \eqref{equ:polarisation_loops} lead to vanishing polarisation loops.

	\clearpage	
	\onecolumngrid
	\section{Parameter sets for different regularisation methods}
	\label{app:parameter_sets}
	\twocolumngrid
	
	\begin{table}[t]
		\caption{\label{tab:parameter_set_4C_eta} Parameter sets with four-momentum cutoff scheme, fit to $\eta$ mass.}
		\begin{ruledtabular}
			\begin{tabular}{ll|lllll}
				&&[C4A]&[C4B]&[C4C]&[C4D]&[C4E]\\
				\hline &&&&&&\\
				$m_u$ & $[\text{MeV}]$ & 275. & 300. & 325. & 350. & 375. \\
				$m_s$ & $[\text{MeV}]$ & 407.059 & 472.862 & 508.499 & 539.649 & 568.744 \\
				$\Lambda$  & $[\text{MeV}]$ & 864.712 & 813.879 & 780.107 & 757.288 & 741.848 \\
				$G\Lambda^2$ &  & 0.434 & 1.742 & 1.986 & 2.141 & 2.274 \\
				$K\Lambda^5$ &  & 341.7 & 234.533 & 234.381 & 245.172 & 259.929 \\
				\hline &&&&&&\\
				$m_{0,u}$ & $[\text{MeV}]$ & 6.529 & 7.19 & 7.682 & 8.038 & 8.286 \\
				$m_{0,s}$ & $[\text{MeV}]$ & 175.06 & 187.221 & 195.661 & 201.353 & 204.973 \\
				$m_{\eta^{\prime}}$ & $[\text{MeV}]$ & 948.594 & 1047.95 & 1090.13 & 1130.31 & 1172.48 \\			
			\end{tabular}
		\end{ruledtabular} \\[1cm]
		\caption{\label{tab:parameter_set_4C_etaprime} Parameter sets with four-momentum cutoff scheme, fit to $\eta^{\prime}$ mass.}
		\begin{ruledtabular}
			\begin{tabular}{ll|lllll}
				&&[C4A']&[C4B']&[C4C']&[C4D']&[C4E']\\
				\hline &&&&&&\\
				$m_u$ & $[\text{MeV}]$ & 275. & 300. & 325. & 350. & 375. \\
				$m_s$ & $[\text{MeV}]$ & 496.977 & 520.78 & 543.293 & 565.059 & 586.331 \\
				$\Lambda$  & $[\text{MeV}]$ & 864.712 & 813.879 & 780.107 & 757.288 & 741.848 \\
				$G\Lambda^2$ &  & 3.487 & 3.711 & 3.946 & 4.195 & 4.458 \\
				$K\Lambda^5$ &  & 63.349 & 66.819 & 71.1 & 75.537 & 79.777 \\
				\hline &&&&&&\\
				$m_{0,u}$ & $[\text{MeV}]$ & 6.529 & 7.19 & 7.682 & 8.038 & 8.286 \\
				$m_{0,s}$ & $[\text{MeV}]$ & 175.765 & 187.475 & 195.645 & 201.178 & 204.741 \\
				$m_{\eta}$ & $[\text{MeV}]$ & 454.361 & 455.479 & 457.737 & 460.34 & 462.866 \\				
			\end{tabular}
		\end{ruledtabular} \\[1cm]
		\caption{\label{tab:parameter_set_4C_etaprime_II} Parameter sets with four-momentum cutoff scheme, fit to $\eta^{\prime}$ mass - \textit{alternative solution}.}
		\begin{ruledtabular}
			\begin{tabular}{ll|lllll}
				&&[C4a']&[C4b']&[C4c']&[C4d']&[C4e']\\
				\hline &&&&&&\\
				$m_u$ & $[\text{MeV}]$ & 275. & 300. & 325. & 350. & 375. \\
				$m_s$ & $[\text{MeV}]$ & 411.648 & 427.756 & 433.564 & 432.229 & 425.302 \\
				$\Lambda$  & $[\text{MeV}]$ & 864.712 & 813.879 & 780.107 & 757.288 & 741.848 \\
				$G\Lambda^2$ &  & 0.642 & -0.646 & -3.661 & -10.55 & -29.71 \\
				$K\Lambda^5$ &  & 321.704 & 447.566 & 723.781 & 1330.18 & 2982.23 \\
				\hline &&&&&&\\
				$m_{0,u}$ & $[\text{MeV}]$ & 6.529 & 7.19 & 7.682 & 8.038 & 8.286 \\
				$m_{0,s}$ & $[\text{MeV}]$ & 175.083 & 186.9 & 195.353 & 201.325 & 205.467 \\
				$m_{\eta}$ & $[\text{MeV}]$ & 547.326 & 562.131 & 569.366 & 573.151 & 574.733 \\				
			\end{tabular}
		\end{ruledtabular}
	\end{table}

	\begin{table}[t]
		\caption{\label{tab:parameter_set_3C_eta} Parameter sets with three-momentum cutoff scheme, fit to $\eta$ mass.}
		\begin{ruledtabular}
			\begin{tabular}{ll|lllll}
				&&[C3A]&[C3B]&[C3C]&[C3D]&[C3E]\\
				\hline &&&&&&\\
				$m_u$ & $[\text{MeV}]$ & 275. & 300. & 325. & 350. & 375. \\
				$m_s$ & $[\text{MeV}]$ & 365.988 & 443.771 & 477.742 & 506.246 & 532.443 \\
				$\Lambda$  & $[\text{MeV}]$ & 705.66 & 661.973 & 632.269 & 611.533 & 596.829 \\
				$G\Lambda^2$ &  & 0.229 & 1.108 & 1.26 & 1.371 & 1.471 \\
				$K\Lambda^5$ &  & 59.27 & 28.085 & 24.873 & 23.472 & 22.635 \\
				\hline &&&&&&\\
				$m_{0,u}$ & $[\text{MeV}]$ & 4.569 & 4.971 & 5.251 & 5.434 & 5.542 \\
				$m_{0,s}$ & $[\text{MeV}]$ & 130.861 & 135.53 & 139.494 & 141.706 & 142.587 \\
				$m_{\eta^{\prime}}$ & $[\text{MeV}]$ & 914.854 & 1091.65 & 1157.84 & 1212.63 & 1269.79 \\		
			\end{tabular}
		\end{ruledtabular} \\[1cm]
		\caption{\label{tab:parameter_set_3C_etaprime} Parameter sets with three-momentum cutoff scheme, fit to $\eta^{\prime}$ mass.}
		\begin{ruledtabular}
			\begin{tabular}{ll|lllll}
				&&[C3A']&[C3B']&[C3C']&[C3D']&[C3E']\\
				\hline &&&&&&\\
				$m_u$ & $[\text{MeV}]$ & 275. & 300. & 325. & 350. & 375. \\
				$m_s$ & $[\text{MeV}]$ & 479.414 & 501.923 & 522.659 & 542.986 & 563.309 \\
				$\Lambda$  & $[\text{MeV}]$ & 705.66 & 661.973 & 632.269 & 611.533 & 596.829 \\
				$G\Lambda^2$ &  & 1.672 & 1.747 & 1.817 & 1.889 & 1.964 \\
				$K\Lambda^5$ &  & 8.793 & 8.809 & 9.109 & 9.439 & 9.714 \\
				\hline &&&&&&\\
				$m_{0,u}$ & $[\text{MeV}]$ & 4.569 & 4.971 & 5.251 & 5.434 & 5.542 \\
				$m_{0,s}$ & $[\text{MeV}]$ & 126.189 & 133.105 & 137.451 & 139.91 & 140.996 \\
				$m_{\eta}$ & $[\text{MeV}]$ & 472.606 & 476.756 & 483.348 & 489.757 & 495.223 \\				
			\end{tabular}
		\end{ruledtabular} \\[1cm]
		\caption{\label{tab:parameter_set_3C_etaprime_II} Parameter sets with three-momentum cutoff scheme, fit to $\eta^{\prime}$ mass - \textit{alternative solution}.}
		\begin{ruledtabular}
			\begin{tabular}{ll|lllll}
				&&[C3a']&[C3b']&[C3c']&[C3d']&[C3e']\\
				\hline &&&&&&\\
				$m_u$ & $[\text{MeV}]$ & 275. & 300. & 325. & 350. & 375. \\
				$m_s$ & $[\text{MeV}]$ & 382.951 & 403.442 & 415.309 & 419.04 & 415.692 \\
				$\Lambda$  & $[\text{MeV}]$ & 705.66 & 661.973 & 632.269 & 611.533 & 596.829 \\
				$G\Lambda^2$ &  & 0.591 & 0.389 & -0.108 & -1.298 & -4.878 \\
				$K\Lambda^5$ &  & 45.679 & 51.758 & 67.236 & 103.062 & 207.964 \\
				\hline &&&&&&\\
				$m_{0,u}$ & $[\text{MeV}]$ & 4.569 & 4.971 & 5.251 & 5.434 & 5.542 \\
				$m_{0,s}$ & $[\text{MeV}]$ & 130.042 & 137.256 & 142.28 & 145.783 & 148.23 \\
				$m_{\eta}$ & $[\text{MeV}]$ & 546.266 & 558.499 & 564.17 & 567.338 & 568.317 \\				
			\end{tabular}
		\end{ruledtabular}
	\end{table}
	
	\begin{table}[t]
		\caption{\label{tab:parameter_set_PV2R_eta} Parameter sets with Pauli-Villars scheme and two regulators, fit to $\eta$ mass.}
		\begin{ruledtabular}
			\begin{tabular}{ll|lllll}
				&&[R2A]&[R2B]&[R2C]&[R2D]&[R2E]\\
				\hline &&&&&&\\
				$m_u$ & $[\text{MeV}]$ & 275. & 300. & 325. & 350. & 375. \\
				$m_s$ & $[\text{MeV}]$ & 404.383 & 470.643 & 506.082 & 536.965 & 565.769 \\
				$\Lambda$  & $[\text{MeV}]$ & 740.399 & 696.497 & 667.227 & 647.354 & 633.813 \\
				$G\Lambda^2$ &  & 0.328 & 1.284 & 1.465 & 1.583 & 1.685 \\
				$K\Lambda^5$ &  & 149.538 & 100.817 & 100.013 & 103.975 & 109.625 \\
				\hline &&&&&&\\
				$m_{0,u}$ & $[\text{MeV}]$ & 6.392 & 7.039 & 7.522 & 7.871 & 8.115 \\
				$m_{0,s}$ & $[\text{MeV}]$ & 172.025 & 183.839 & 192.117 & 197.727 & 201.322 \\
				$m_{\eta^{\prime}}$ & $[\text{MeV}]$ & 945.778 & 1047.06 & 1088.62 & 1127.75 & 1168.78 \\	
			\end{tabular}
		\end{ruledtabular} \\[1cm]
		\caption{\label{tab:parameter_set_PV2R_etaprime} Parameter sets with Pauli-Villars scheme and two regulators, fit to $\eta^{\prime}$ mass.}
		\begin{ruledtabular}
			\begin{tabular}{ll|lllll}
				&&[R2A']&[R2B']&[R2C']&[R2D']&[R2E']\\
				\hline &&&&&&\\
				$m_u$ & $[\text{MeV}]$ & 275. & 300. & 325. & 350. & 375. \\
				$m_s$ & $[\text{MeV}]$ & 495.89 & 519.597 & 542.035 & 563.763 & 585.036 \\
				$\Lambda$  & $[\text{MeV}]$ & 740.399 & 696.497 & 667.227 & 647.354 & 633.813 \\
				$G\Lambda^2$ &  & 2.509 & 2.666 & 2.831 & 3.005 & 3.189 \\
				$K\Lambda^5$ &  & 27.46 & 28.907 & 30.728 & 32.623 & 34.429 \\
				\hline &&&&&&\\
				$m_{0,u}$ & $[\text{MeV}]$ & 6.392 & 7.039 & 7.522 & 7.871 & 8.115 \\
				$m_{0,s}$ & $[\text{MeV}]$ & 172.467 & 183.981 & 192.033 & 197.507 & 201.053 \\
				$m_{\eta}$ & $[\text{MeV}]$ & 455.349 & 456.763 & 459.339 & 462.211 & 464.925 \\			
			\end{tabular}
		\end{ruledtabular} \\[1cm]
		\caption{\label{tab:parameter_set_PV2R_etaprime_II} Parameter sets with Pauli-Villars scheme and two regulators, fit to $\eta^{\prime}$ mass - \textit{alternative solution}.}
		\begin{ruledtabular}
			\begin{tabular}{ll|lllll}
				&&[R2a']&[R2b']&[R2c']&[R2d']&[R2e']\\
				\hline &&&&&&\\
				$m_u$ & $[\text{MeV}]$ & 275. & 300. & 325. & 350. & 375. \\
				$m_s$ & $[\text{MeV}]$ & 410.266 & 426.897 & 433.171 & 432.143 & 425.436 \\
				$\Lambda$  & $[\text{MeV}]$ & 740.399 & 696.497 & 667.227 & 647.354 & 633.813 \\
				$G\Lambda^2$ &  & 0.517 & -0.33 & -2.319 & -6.855 & -19.41 \\
				$K\Lambda^5$ &  & 138.336 & 188.983 & 300.458 & 544.345 & 1204.2 \\
				\hline &&&&&&\\
				$m_{0,u}$ & $[\text{MeV}]$ & 6.392 & 7.039 & 7.522 & 7.871 & 8.115 \\
				$m_{0,s}$ & $[\text{MeV}]$ & 172.035 & 183.653 & 191.997 & 197.919 & 202.053 \\
				$m_{\eta}$ & $[\text{MeV}]$ & 547.139 & 561.705 & 568.918 & 572.752 & 574.357 \\			
			\end{tabular}
		\end{ruledtabular}
	\end{table}
	
	\begin{table}[t]
		\caption{\label{tab:parameter_set_PV3R_eta} Parameter sets with Pauli-Villars scheme and three regulators, fit to $\eta$ mass.}
		\begin{ruledtabular}
			\begin{tabular}{ll|lllll}
				&&[R3A]&[R3B]&[R3C]&[R3D]&[R3E]\\
				\hline &&&&&&\\
				$m_u$ & $[\text{MeV}]$ & 275. & 300. & 325. & 350. & 375. \\
				$m_s$ & $[\text{MeV}]$ & 435.703 & 504.503 & 553.724 & 618.265 & \\
				$\Lambda$  & $[\text{MeV}]$ & 862.033 & 813.116 & 781.181 & 760.167 & \\
				$G\Lambda^2$ &  & -0.119 & 1.093 & 0.867 & -0.197 & \\
				$K\Lambda^5$ &  & 603.131 & 512.35 & 606.471 & 833.135 & \\
				\hline &&&&&&\\
				$m_{0,u}$ & $[\text{MeV}]$ & 8.002 & 8.896 & 9.596 & 10.134 & \\
				$m_{0,s}$ & $[\text{MeV}]$ & 208.388 & 227.383 & 241.122 & 252.073 & \\
				$m_{\eta^{\prime}}$ & $[\text{MeV}]$ & 973.151 & 1067.06 & 1134.83 & 1240.75 & \\
			\end{tabular}
		\end{ruledtabular} \\[1cm]
		\caption{\label{tab:parameter_set_PV3R_etaprime} Parameter sets with Pauli-Villars scheme and three regulators, fit to $\eta^{\prime}$ mass.}
		\begin{ruledtabular}
			\begin{tabular}{ll|lllll}
				&&[R3A']&[R3B']&[R3C']&[R3D']&[R3E']\\
				\hline &&&&&&\\
				$m_u$ & $[\text{MeV}]$ & 275. & 300. & 325. & 350. & 375. \\
				$m_s$ & $[\text{MeV}]$ & 509.785 & 535.055 & 558.779 & 581.39 & 603.149 \\
				$\Lambda$  & $[\text{MeV}]$ & 862.033 & 813.116 & 781.181 & 760.167 & 746.523 \\
				$G\Lambda^2$ &  & 4.203 & 4.551 & 4.928 & 5.332 & 5.762 \\
				$K\Lambda^5$ &  & 108.561 & 117.963 & 128.717 & 140.057 & 151.561 \\
				\hline &&&&&&\\
				$m_{0,u}$ & $[\text{MeV}]$ & 8.002 & 8.896 & 9.596 & 10.134 & 10.54 \\
				$m_{0,s}$ & $[\text{MeV}]$ & 212.992 & 229.28 & 241.406 & 250.368 & 256.913 \\
				$m_{\eta}$ & $[\text{MeV}]$ & 437.87 & 434.372 & 432.23 & 431.025 & 430.495 \\			
			\end{tabular}
		\end{ruledtabular} \\[1cm]
		\caption{\label{tab:parameter_set_PV3R_etaprime_II} Parameter sets with Pauli-Villars scheme and three regulators, fit to $\eta^{\prime}$ mass - \textit{alternative solution}.}
		\begin{ruledtabular}
			\begin{tabular}{ll|lllll}
				&&[R3a']&[R3b']&[R3c']&[R3d']&[R3e']\\
				\hline &&&&&&\\
				$m_u$ & $[\text{MeV}]$ & 275. & 300. & 325. & 350. & 375. \\
				$m_s$ & $[\text{MeV}]$ & 427.127 & 438.532 & 440.748 & 437.314 & 429.101 \\
				$\Lambda$  & $[\text{MeV}]$ & 862.033 & 813.116 & 781.181 & 760.167 & 746.523 \\
				$G\Lambda^2$ &  & -0.675 & -5.332 & -18.64 & -62.18 & -270.6 \\
				$K\Lambda^5$ &  & 668.388 & 1240.22 & 2802.28 & 7807.21 & 31557.3 \\
				\hline &&&&&&\\
				$m_{0,u}$ & $[\text{MeV}]$ & 8.002 & 8.896 & 9.596 & 10.134 & 10.54 \\
				$m_{0,s}$ & $[\text{MeV}]$ & 207.875 & 223.133 & 234.067 & 241.831 & 247.183 \\
				$m_{\eta}$ & $[\text{MeV}]$ & 549.163 & 566.818 & 574.267 & 577.898 & 579.853 \\			
			\end{tabular}
		\end{ruledtabular}
	\end{table}
	

	\clearpage
	
	\onecolumngrid	
	\section{Comparison between real part and full integrals}
	\label{app:real_vs_full_integral}

	\begin{table*}[h]
		\caption{\label{tab:full_vs_real_integral_PV2R_etaprime} Comparing parameter sets with full and only real part of integrals as function of $m_s$ (Pauli-Villars scheme with two regulators fit to $\eta^{\prime}$ mass).}
		\begin{ruledtabular}
			\begin{tabular}{ll|llllll>{\columncolor{green!15}[0pt]}l>{\columncolor{green!15}[0pt]}l>{\columncolor{green!15}[0pt]}l>{\columncolor{green!15}[0pt]}l>{\columncolor{green!15}[0pt]}l}
				\rowcolor{white}$m_s$ & $[\text{MeV}]$ & 430. & 460. & 470. & 476. & 477. & 478. & 479. & 480. & 490. & 500. & 530. \\
				\rowcolor{white} \hline &&&&&&&&&&&&\\
				\rowcolor{white}\multicolumn{2}{l|}{\textit{full integral}} &&&&&&&&&&&\\[0.1cm]
				$m_u$ & $[\text{MeV}]$ & 176.219 & 210.395 & 230.203 & 246.241 & 249.428 & 253.003 & 259.023 & 259.704 & 269.117 & 279.196 & 311.465 \\
				$\Lambda$  & $[\text{MeV}]$ & 1496.79 & 1021.18 & 891.821 & 821.866 & 810.5 & 798.57 & 780.238 & 778.294 & 753.755 & 731.677 & 681.641 \\
				$G\Lambda^2$ &  & 2.369 & 2.467 & 2.493 & 2.472 & 2.462 & 2.446 & 2.404 & 2.412 & 2.473 & 2.535 & 2.74 \\
				$K\Lambda^5$ &  & 7.186 & 8.589 & 13.047 & 19.36 & 20.996 & 23.047 & 27.413 & 27.196 & 27.237 & 27.657 & 29.717 \\[0.1cm]
				$m_{0,u}$ & $[\text{MeV}]$ & 1.987 & 3.786 & 4.733 & 5.406 & 5.529 & 5.663 & 5.879 & 5.902 & 6.213 & 6.514 & 7.279 \\
				$m_{0,s}$ & $[\text{MeV}]$ & 70.834 & 118.212 & 139.549 & 153.507 & 155.961 & 158.598 & 162.769 & 163.227 & 169.135 & 174.688 & 188.044 \\
				$m_{\eta}$ & $[\text{MeV}]$ & 330.935 & 334.019 & 375.166 & 418.973 & 428.076 & 438.446 & 457.239 & 456.316 & 455.34 & 455.454 & 457.861 \\	
				\rowcolor{white}\hline &&&&&&&&&&&&\\
				\rowcolor{white}\multicolumn{2}{l|}{\textit{only real part}} &&&&&&&&&&&\\[0.1cm]
				$m_u$ & $[\text{MeV}]$ & 213.294 & 240.232 & 249.958 & 255.969 & 256.984 & 258.002 & 259.023 & 259.704 & 269.117 & 279.196 & 311.465 \\
				$\Lambda$  & $[\text{MeV}]$ & 998.192 & 845.377 & 808.68 & 789.275 & 786.215 & 783.203 & 780.238 & 778.294 & 753.755 & 731.677 & 681.641 \\
				$G\Lambda^2$ &  & 2.227 & 2.322 & 2.363 & 2.39 & 2.394 & 2.399 & 2.404 & 2.412 & 2.473 & 2.535 & 2.74 \\
				$K\Lambda^5$ &  & 24.532 & 25.888 & 26.616 & 27.132 & 27.224 & 27.318 & 27.413 & 27.196 & 27.237 & 27.657 & 29.717 \\[0.1cm]
				$m_{0,u}$ & $[\text{MeV}]$ & 3.932 & 5.164 & 5.549 & 5.771 & 5.807 & 5.843 & 5.879 & 5.902 & 6.213 & 6.514 & 7.279 \\
				$m_{0,s}$ & $[\text{MeV}]$ & 121.503 & 148.491 & 156.303 & 160.674 & 161.379 & 162.078 & 162.769 & 163.227 & 169.135 & 174.688 & 188.044 \\
				$m_{\eta}$ & $[\text{MeV}]$ & 436.972 & 451.18 & 454.521 & 456.342 & 456.641 & 456.94 & 457.239 & 456.316 & 455.34 & 455.454 & 457.861 \\	
			\end{tabular}
		\end{ruledtabular}
	\end{table*}

	\twocolumngrid
	

	\bibliography{literature}

\begin{thebibliography}{24}%
\makeatletter
\providecommand \@ifxundefined [1]{%
 \@ifx{#1\undefined}
}%
\providecommand \@ifnum [1]{%
 \ifnum #1\expandafter \@firstoftwo
 \else \expandafter \@secondoftwo
 \fi
}%
\providecommand \@ifx [1]{%
 \ifx #1\expandafter \@firstoftwo
 \else \expandafter \@secondoftwo
 \fi
}%
\providecommand \natexlab [1]{#1}%
\providecommand \enquote  [1]{``#1''}%
\providecommand \bibnamefont  [1]{#1}%
\providecommand \bibfnamefont [1]{#1}%
\providecommand \citenamefont [1]{#1}%
\providecommand \href@noop [0]{\@secondoftwo}%
\providecommand \href [0]{\begingroup \@sanitize@url \@href}%
\providecommand \@href[1]{\@@startlink{#1}\@@href}%
\providecommand \@@href[1]{\endgroup#1\@@endlink}%
\providecommand \@sanitize@url [0]{\catcode `\\12\catcode `\$12\catcode
  `\&12\catcode `\#12\catcode `\^12\catcode `\_12\catcode `\%12\relax}%
\providecommand \@@startlink[1]{}%
\providecommand \@@endlink[0]{}%
\providecommand \url  [0]{\begingroup\@sanitize@url \@url }%
\providecommand \@url [1]{\endgroup\@href {#1}{\urlprefix }}%
\providecommand \urlprefix  [0]{URL }%
\providecommand \Eprint [0]{\href }%
\providecommand \doibase [0]{https://doi.org/}%
\providecommand \selectlanguage [0]{\@gobble}%
\providecommand \bibinfo  [0]{\@secondoftwo}%
\providecommand \bibfield  [0]{\@secondoftwo}%
\providecommand \translation [1]{[#1]}%
\providecommand \BibitemOpen [0]{}%
\providecommand \bibitemStop [0]{}%
\providecommand \bibitemNoStop [0]{.\EOS\space}%
\providecommand \EOS [0]{\spacefactor3000\relax}%
\providecommand \BibitemShut  [1]{\csname bibitem#1\endcsname}%
\let\auto@bib@innerbib\@empty
\bibitem [{\citenamefont {Nambu}\ and\ \citenamefont
  {Jona-Lasinio}(1961{\natexlab{a}})}]{PhysRev.122.345}%
  \BibitemOpen
  \bibfield  {author} {\bibinfo {author} {\bibfnamefont {Y.}~\bibnamefont
  {Nambu}}\ and\ \bibinfo {author} {\bibfnamefont {G.}~\bibnamefont
  {Jona-Lasinio}},\ }\bibfield  {title} {\bibinfo {title} {Dynamical model of
  elementary particles based on an analogy with superconductivity. i},\ }\href
  {https://doi.org/10.1103/PhysRev.122.345} {\bibfield  {journal} {\bibinfo
  {journal} {Phys. Rev.}\ }\textbf {\bibinfo {volume} {122}},\ \bibinfo {pages}
  {345} (\bibinfo {year} {1961}{\natexlab{a}})}\BibitemShut {NoStop}%
\bibitem [{\citenamefont {Nambu}\ and\ \citenamefont
  {Jona-Lasinio}(1961{\natexlab{b}})}]{PhysRev.124.246}%
  \BibitemOpen
  \bibfield  {author} {\bibinfo {author} {\bibfnamefont {Y.}~\bibnamefont
  {Nambu}}\ and\ \bibinfo {author} {\bibfnamefont {G.}~\bibnamefont
  {Jona-Lasinio}},\ }\bibfield  {title} {\bibinfo {title} {Dynamical model of
  elementary particles based on an analogy with superconductivity. ii},\ }\href
  {https://doi.org/10.1103/PhysRev.124.246} {\bibfield  {journal} {\bibinfo
  {journal} {Phys. Rev.}\ }\textbf {\bibinfo {volume} {124}},\ \bibinfo {pages}
  {246} (\bibinfo {year} {1961}{\natexlab{b}})}\BibitemShut {NoStop}%
\bibitem [{\citenamefont {Buballa}(2005)}]{njl-buballa}%
  \BibitemOpen
  \bibfield  {author} {\bibinfo {author} {\bibfnamefont {M.}~\bibnamefont
  {Buballa}},\ }\bibfield  {title} {\bibinfo {title} {{NJL-model analysis of
  dense quark matter}},\ }\href {https://doi.org/10.1016/j.physrep.2004.11.004}
  {\bibfield  {journal} {\bibinfo  {journal} {Phys. Rept.}\ }\textbf {\bibinfo
  {volume} {407(4-6)}},\ \bibinfo {pages} {205} (\bibinfo {year} {2005})},\
  \Eprint {https://arxiv.org/abs/hep-ph/0402234} {arXiv:hep-ph/0402234}
  \BibitemShut {NoStop}%
\bibitem [{\citenamefont {Oertel}(2000)}]{oertel2000investigation}%
  \BibitemOpen
  \bibfield  {author} {\bibinfo {author} {\bibfnamefont {M.}~\bibnamefont
  {Oertel}},\ }\href@noop {} {\bibinfo {type} {{PhD-Thesis}}},\ \bibinfo
  {school} {Technische Universit\"at Darmstadt} (\bibinfo {year} {2000}),\
  \Eprint {https://arxiv.org/abs/hep-ph/0012224} {arXiv:hep-ph/0012224}
  \BibitemShut {NoStop}%
\bibitem [{\citenamefont {Braun-Munzinger}\ and\ \citenamefont
  {Wambach}(2009)}]{BraunMunzinger:2008tz}%
  \BibitemOpen
  \bibfield  {author} {\bibinfo {author} {\bibfnamefont {P.}~\bibnamefont
  {Braun-Munzinger}}\ and\ \bibinfo {author} {\bibfnamefont {J.}~\bibnamefont
  {Wambach}},\ }\bibfield  {title} {\bibinfo {title} {{The Phase Diagram of
  Strongly-Interacting Matter}},\ }\href
  {https://doi.org/10.1103/RevModPhys.81.1031} {\bibfield  {journal} {\bibinfo
  {journal} {Rev. Mod. Phys.}\ }\textbf {\bibinfo {volume} {81(3)}},\ \bibinfo
  {pages} {1031} (\bibinfo {year} {2009})},\ \Eprint
  {https://arxiv.org/abs/0801.4256} {arXiv:0801.4256 [hep-ph]} \BibitemShut
  {NoStop}%
\bibitem [{\citenamefont {Klevansky}(1992)}]{njl-klevansky}%
  \BibitemOpen
  \bibfield  {author} {\bibinfo {author} {\bibfnamefont {S.}~\bibnamefont
  {Klevansky}},\ }\bibfield  {title} {\bibinfo {title} {{The
  Nambu--Jona-Lasinio model of quantum chromodynamics}},\ }\href
  {https://doi.org/10.1103/RevModPhys.64.649} {\bibfield  {journal} {\bibinfo
  {journal} {Rev. Mod. Phys.}\ }\textbf {\bibinfo {volume} {64(3)}},\ \bibinfo
  {pages} {649} (\bibinfo {year} {1992})}\BibitemShut {NoStop}%
\bibitem [{\citenamefont {A.~Buck}\ and\ \citenamefont
  {Reinhardt}(1992)}]{Buck:1992}%
  \BibitemOpen
  \bibfield  {author} {\bibinfo {author} {\bibfnamefont {R.~A.}\ \bibnamefont
  {A.~Buck}}\ and\ \bibinfo {author} {\bibfnamefont {H.}~\bibnamefont
  {Reinhardt}},\ }\bibfield  {title} {\bibinfo {title} {{Baryons as bound
  states of diquarks and quarks in the Nambu--Jona-Lasinio model}},\ }\href
  {https://doi.org/10.1016/0370-2693(92)90154-V} {\bibfield  {journal}
  {\bibinfo  {journal} {Phys. Lett.}\ }\textbf {\bibinfo {volume} {B268}},\
  \bibinfo {pages} {29} (\bibinfo {year} {1992})}\BibitemShut {NoStop}%
\bibitem [{\citenamefont {Torres-Rincon}\ \emph {et~al.}(2015)\citenamefont
  {Torres-Rincon}, \citenamefont {Sintes},\ and\ \citenamefont
  {Aichelin}}]{Torres-Rincon:2015rma}%
  \BibitemOpen
  \bibfield  {author} {\bibinfo {author} {\bibfnamefont {J.~M.}\ \bibnamefont
  {Torres-Rincon}}, \bibinfo {author} {\bibfnamefont {B.}~\bibnamefont
  {Sintes}},\ and\ \bibinfo {author} {\bibfnamefont {J.}~\bibnamefont
  {Aichelin}},\ }\bibfield  {title} {\bibinfo {title} {{Flavor dependence of
  baryon melting temperature in effective models of QCD}},\ }\href
  {https://doi.org/10.1103/PhysRevC.91.065206} {\bibfield  {journal} {\bibinfo
  {journal} {Phys. Rev.}\ }\textbf {\bibinfo {volume} {C91}},\ \bibinfo {pages}
  {065206} (\bibinfo {year} {2015})},\ \Eprint
  {https://arxiv.org/abs/1502.03459} {arXiv:1502.03459 [hep-ph]} \BibitemShut
  {NoStop}%
\bibitem [{\citenamefont {Kohyama}\ \emph {et~al.}(2016)\citenamefont
  {Kohyama}, \citenamefont {Kimura},\ and\ \citenamefont
  {Inagaki}}]{Kohyama:2016fif}%
  \BibitemOpen
  \bibfield  {author} {\bibinfo {author} {\bibfnamefont {H.}~\bibnamefont
  {Kohyama}}, \bibinfo {author} {\bibfnamefont {D.}~\bibnamefont {Kimura}},\
  and\ \bibinfo {author} {\bibfnamefont {T.}~\bibnamefont {Inagaki}},\
  }\bibfield  {title} {\bibinfo {title} {{Parameter fitting in three-flavor
  Nambu--Jona-Lasinio model with various regularizations}},\ }\href
  {https://doi.org/10.1016/j.nuclphysb.2016.03.015} {\bibfield  {journal}
  {\bibinfo  {journal} {Nucl. Phys. B}\ }\textbf {\bibinfo {volume} {906}},\
  \bibinfo {pages} {524} (\bibinfo {year} {2016})},\ \Eprint
  {https://arxiv.org/abs/1601.02411} {arXiv:1601.02411 [hep-ph]} \BibitemShut
  {NoStop}%
\bibitem [{\citenamefont {Ratti}\ \emph {et~al.}(2006)\citenamefont {Ratti},
  \citenamefont {Thaler},\ and\ \citenamefont {Weise}}]{PhysRevD.73.014019}%
  \BibitemOpen
  \bibfield  {author} {\bibinfo {author} {\bibfnamefont {C.}~\bibnamefont
  {Ratti}}, \bibinfo {author} {\bibfnamefont {M.~A.}\ \bibnamefont {Thaler}},\
  and\ \bibinfo {author} {\bibfnamefont {W.}~\bibnamefont {Weise}},\ }\bibfield
   {title} {\bibinfo {title} {Phases of qcd: Lattice thermodynamics and a field
  theoretical model},\ }\href {https://doi.org/10.1103/PhysRevD.73.014019}
  {\bibfield  {journal} {\bibinfo  {journal} {Phys. Rev. D}\ }\textbf {\bibinfo
  {volume} {73}},\ \bibinfo {pages} {014019} (\bibinfo {year}
  {2006})}\BibitemShut {NoStop}%
\bibitem [{\citenamefont {Blanquier}(2017)}]{Blanquier:2016dls}%
  \BibitemOpen
  \bibfield  {author} {\bibinfo {author} {\bibfnamefont {E.}~\bibnamefont
  {Blanquier}},\ }\bibfield  {title} {\bibinfo {title} {{Color
  superconductivity in the Nambu-Jona-Lasinio model complemented by a Polyakov
  loop}},\ }\href {https://doi.org/10.1140/epja/i2017-12317-3} {\bibfield
  {journal} {\bibinfo  {journal} {Eur. Phys. J. A}\ }\textbf {\bibinfo {volume}
  {53}},\ \bibinfo {pages} {137} (\bibinfo {year} {2017})},\ \Eprint
  {https://arxiv.org/abs/1606.02672} {arXiv:1606.02672 [hep-ph]} \BibitemShut
  {NoStop}%
\bibitem [{\citenamefont {Hatsuda}\ and\ \citenamefont
  {Kunihiro}(1994)}]{HATSUDA1994221}%
  \BibitemOpen
  \bibfield  {author} {\bibinfo {author} {\bibfnamefont {T.}~\bibnamefont
  {Hatsuda}}\ and\ \bibinfo {author} {\bibfnamefont {T.}~\bibnamefont
  {Kunihiro}},\ }\bibfield  {title} {\bibinfo {title} {Qcd phenomenology based
  on a chiral effective lagrangian},\ }\href
  {https://doi.org/10.1016/0370-1573(94)90022-1} {\bibfield  {journal}
  {\bibinfo  {journal} {Physics Reports}\ }\textbf {\bibinfo {volume} {247}},\
  \bibinfo {pages} {221} (\bibinfo {year} {1994})}\BibitemShut {NoStop}%
\bibitem [{\citenamefont {Rehberg}\ \emph {et~al.}(1996)\citenamefont
  {Rehberg}, \citenamefont {Klevansky},\ and\ \citenamefont
  {Hufner}}]{Rehberg:1995kh}%
  \BibitemOpen
  \bibfield  {author} {\bibinfo {author} {\bibfnamefont {P.}~\bibnamefont
  {Rehberg}}, \bibinfo {author} {\bibfnamefont {S.}~\bibnamefont {Klevansky}},\
  and\ \bibinfo {author} {\bibfnamefont {J.}~\bibnamefont {Hufner}},\
  }\bibfield  {title} {\bibinfo {title} {{Hadronization in the SU(3)
  Nambu-Jona-Lasinio model}},\ }\href {https://doi.org/10.1103/PhysRevC.53.410}
  {\bibfield  {journal} {\bibinfo  {journal} {Phys. Rev. C}\ }\textbf {\bibinfo
  {volume} {53}},\ \bibinfo {pages} {410} (\bibinfo {year} {1996})},\ \Eprint
  {https://arxiv.org/abs/hep-ph/9506436} {arXiv:hep-ph/9506436} \BibitemShut
  {NoStop}%
\bibitem [{\citenamefont {Klimt}\ \emph {et~al.}(1990)\citenamefont {Klimt},
  \citenamefont {Lutz}, \citenamefont {Vogl},\ and\ \citenamefont
  {Weise}}]{Klimt:1989pm}%
  \BibitemOpen
  \bibfield  {author} {\bibinfo {author} {\bibfnamefont {S.}~\bibnamefont
  {Klimt}}, \bibinfo {author} {\bibfnamefont {M.~F.}\ \bibnamefont {Lutz}},
  \bibinfo {author} {\bibfnamefont {U.}~\bibnamefont {Vogl}},\ and\ \bibinfo
  {author} {\bibfnamefont {W.}~\bibnamefont {Weise}},\ }\bibfield  {title}
  {\bibinfo {title} {{GENERALIZED SU(3) NAMBU-JONA-LASINIO MODEL. Part. 1.
  MESONIC MODES}},\ }\href {https://doi.org/10.1016/0375-9474(90)90123-4}
  {\bibfield  {journal} {\bibinfo  {journal} {Nucl. Phys. A}\ }\textbf
  {\bibinfo {volume} {516}},\ \bibinfo {pages} {429} (\bibinfo {year}
  {1990})}\BibitemShut {NoStop}%
\bibitem [{\citenamefont {Vogl}\ \emph {et~al.}(1990)\citenamefont {Vogl},
  \citenamefont {Lutz}, \citenamefont {Klimt},\ and\ \citenamefont
  {Weise}}]{Vogl:1989ea}%
  \BibitemOpen
  \bibfield  {author} {\bibinfo {author} {\bibfnamefont {U.}~\bibnamefont
  {Vogl}}, \bibinfo {author} {\bibfnamefont {M.~F.}\ \bibnamefont {Lutz}},
  \bibinfo {author} {\bibfnamefont {S.}~\bibnamefont {Klimt}},\ and\ \bibinfo
  {author} {\bibfnamefont {W.}~\bibnamefont {Weise}},\ }\bibfield  {title}
  {\bibinfo {title} {{Generalized SU(3) "Nambu-Jona-Lasinio" Model. Part 2.
  From Current to Constituent Quarks}},\ }\href
  {https://doi.org/10.1016/0375-9474(90)90124-5} {\bibfield  {journal}
  {\bibinfo  {journal} {Nucl. Phys. A}\ }\textbf {\bibinfo {volume} {516}},\
  \bibinfo {pages} {469} (\bibinfo {year} {1990})}\BibitemShut {NoStop}%
\bibitem [{\citenamefont {Alkofer}\ and\ \citenamefont
  {Reinhardt}(1995)}]{Alkofer:295076}%
  \BibitemOpen
  \bibfield  {author} {\bibinfo {author} {\bibfnamefont {R.}~\bibnamefont
  {Alkofer}}\ and\ \bibinfo {author} {\bibfnamefont {H.}~\bibnamefont
  {Reinhardt}},\ }\href {https://cds.cern.ch/record/295076} {\emph {\bibinfo
  {title} {{Chiral quark dynamics}}}},\ Lecture Notes in Physics Monographs\
  (\bibinfo  {publisher} {Springer},\ \bibinfo {address} {Berlin},\ \bibinfo
  {year} {1995})\BibitemShut {NoStop}%
\bibitem [{\citenamefont {Reinhardt}(1990)}]{Reinhardt:1989rw}%
  \BibitemOpen
  \bibfield  {author} {\bibinfo {author} {\bibfnamefont {H.}~\bibnamefont
  {Reinhardt}},\ }\bibfield  {title} {\bibinfo {title} {{Hadronization of Quark
  Flavor Dynamics}},\ }\href {https://doi.org/10.1016/0370-2693(90)90078-K}
  {\bibfield  {journal} {\bibinfo  {journal} {Phys. Lett.}\ }\textbf {\bibinfo
  {volume} {B244}},\ \bibinfo {pages} {316} (\bibinfo {year}
  {1990})}\BibitemShut {NoStop}%
\bibitem [{\citenamefont {Osipov}\ \emph {et~al.}(2004)\citenamefont {Osipov},
  \citenamefont {Blin},\ and\ \citenamefont {Hiller}}]{Osipov:2004mn}%
  \BibitemOpen
  \bibfield  {author} {\bibinfo {author} {\bibfnamefont {A.~A.}\ \bibnamefont
  {Osipov}}, \bibinfo {author} {\bibfnamefont {A.~H.}\ \bibnamefont {Blin}},\
  and\ \bibinfo {author} {\bibfnamefont {B.}~\bibnamefont {Hiller}},\
  }\bibfield  {title} {\bibinfo {title} {{The 't Hooft determinant resolution
  of the eta-prime puzzle}},\ }\href@noop {} {\  (\bibinfo {year} {2004})},\
  \Eprint {https://arxiv.org/abs/hep-ph/0410148} {arXiv:hep-ph/0410148}
  \BibitemShut {NoStop}%
\bibitem [{\citenamefont {Burgess}\ and\ \citenamefont
  {Moore}(2007)}]{burgess2007standard}%
  \BibitemOpen
  \bibfield  {author} {\bibinfo {author} {\bibfnamefont {C.}~\bibnamefont
  {Burgess}}\ and\ \bibinfo {author} {\bibfnamefont {G.}~\bibnamefont
  {Moore}},\ }\href {https://doi.org/10.1017/CBO9780511819698} {\emph {\bibinfo
  {title} {The Standard Model: A Primer}}},\ Cambridge books online\ (\bibinfo
  {publisher} {Cambridge University Press},\ \bibinfo {year}
  {2007})\BibitemShut {NoStop}%
\bibitem [{\citenamefont {Peskin}\ and\ \citenamefont
  {Schroeder}(1995)}]{Peskin1995}%
  \BibitemOpen
  \bibfield  {author} {\bibinfo {author} {\bibfnamefont {M.~E.}\ \bibnamefont
  {Peskin}}\ and\ \bibinfo {author} {\bibfnamefont {D.~V.}\ \bibnamefont
  {Schroeder}},\ }\href@noop {} {\emph {\bibinfo {title} {An Introduction To
  Quantum Field Theory}}}\ (\bibinfo  {publisher} {Westview Press},\ \bibinfo
  {address} {New York},\ \bibinfo {year} {1995})\BibitemShut {NoStop}%
\bibitem [{\citenamefont {Pauli}\ and\ \citenamefont
  {Villars}(1949)}]{pauli1949invariant}%
  \BibitemOpen
  \bibfield  {author} {\bibinfo {author} {\bibfnamefont {W.}~\bibnamefont
  {Pauli}}\ and\ \bibinfo {author} {\bibfnamefont {F.}~\bibnamefont
  {Villars}},\ }\bibfield  {title} {\bibinfo {title} {{On the Invariant
  Regularization in Relativistic Quantum Theory}},\ }\href
  {https://doi.org/10.1103/RevModPhys.21.434} {\bibfield  {journal} {\bibinfo
  {journal} {Rev. Mod. Phys.}\ }\textbf {\bibinfo {volume} {21(3)}},\ \bibinfo
  {pages} {434} (\bibinfo {year} {1949})}\BibitemShut {NoStop}%
\bibitem [{\citenamefont {M\"oller}(2012)}]{moeller-master}%
  \BibitemOpen
  \bibfield  {author} {\bibinfo {author} {\bibfnamefont {S.}~\bibnamefont
  {M\"oller}},\ }\href
  {https://theorie.ikp.physik.tu-darmstadt.de/nhq/downloads/thesis/master.moeller.pdf}
  {\bibinfo {type} {{Master-Thesis}}},\ \bibinfo  {school} {Technische
  Universit\"at Darmstadt} (\bibinfo {year} {2012})\BibitemShut {NoStop}%
\bibitem [{\citenamefont {Ossola}\ and\ \citenamefont
  {Sirlin}(2003)}]{Ossola:2003ku}%
  \BibitemOpen
  \bibfield  {author} {\bibinfo {author} {\bibfnamefont {G.}~\bibnamefont
  {Ossola}}\ and\ \bibinfo {author} {\bibfnamefont {A.}~\bibnamefont
  {Sirlin}},\ }\bibfield  {title} {\bibinfo {title} {{Considerations concerning
  the contributions of fundamental particles to the vacuum energy density}},\
  }\href {https://doi.org/10.1140/epjc/s2003-01337-7} {\bibfield  {journal}
  {\bibinfo  {journal} {Eur. Phys. J. C}\ }\textbf {\bibinfo {volume} {31}},\
  \bibinfo {pages} {165} (\bibinfo {year} {2003})},\ \Eprint
  {https://arxiv.org/abs/hep-ph/0305050} {arXiv:hep-ph/0305050} \BibitemShut
  {NoStop}%
\bibitem [{\citenamefont {{Wolfram Research, Inc.}}(2019)}]{Mathematica}%
  \BibitemOpen
  \bibfield  {author} {\bibinfo {author} {\bibnamefont {{Wolfram Research,
  Inc.}}},\ }\href {https://www.wolfram.com/mathematica/} {\bibinfo {title}
  {{Mathematica 12}}} (\bibinfo {year} {2019})\BibitemShut {NoStop}%
\end{thebibliography}%
	
\end{document}